\documentclass[traditabstract]{aa}

\pdfoutput=1
\usepackage{graphicx}
\usepackage[font=small]{caption}
\usepackage{indentfirst}
\usepackage{tabu}
\usepackage[percent]{overpic}
\usepackage{transparent}
\usepackage{ragged2e}
\usepackage[utf8]{inputenc}

\usepackage{natbib}
\usepackage{booktabs}
\usepackage{fixltx2e}

\usepackage{url}
\usepackage{hyperref}

\usepackage{amssymb}
\usepackage{txfonts}
\usepackage[a]{esvect}
\usepackage{tikz}

\newcommand{\degree}[0]{$^{\circ}$}

\let\baraccent=\=
\renewcommand{\=}[1]{\stackrel{#1}{=}} 
\let\arrowaccent=\>
\renewcommand{\>}[1]{\stackrel{#1}{\Rightarrow}} 

\makeatletter
\newcommand{\rmnum}[1]{{\footnotesize{\expandafter\@slowromancap\romannumeral #1@}}}
\newcommand{\Rmnum}[1]{{\expandafter\@slowromancap\romannumeral #1@}}
\makeatother

\everymath={\rm}

\newcommand{\CI}{$[$C\rmnum{1}$]$}

\hypersetup{
   colorlinks=true,       
   linkcolor=black,       
   citecolor=blue,        
   urlcolor=black         
}

\title{Physical conditions in Centaurus A's northern filaments \Rmnum{1}: \\ APEX mid-J CO observations of CO-bright regions\thanks{This publication is based on data acquired with the Atacama Pathfinder Experiment (APEX) under programme ID 098.B-0004, and with the Atacama Large Millimeter/submillimeter Array (ALMA) under programme ADS/JAO.ALMA$\#$2015.1.01019.S.}$^,$\thanks{The reduced APEX spectra (FITS files) are only available at the CDS via anonymous ftp to cdsarc.u-strasbg.fr (130.79.128.5) or via http://cdsweb.u-strasbg.fr/cgi-bin/qcat?J/A+A/.}}

\author{
   Q. Salom\'e$^{1}$ \and
   P. Salom\'e$^{2}$ \and
   A. Gusdorf$^{2}$  \and
   F. Combes$^{2,3}$
}

\institute{
   Instituto de Radioastronom\'ia y Astrof\'isica, Universidad Nacional Aut\'onoma de M\'exico, 58089 Morelia, Michoac\'an, M\'exico \\ email: q.salome@irya.unam.mx \and
   LERMA, Observatoire de Paris, Ecole Normale Supérieure, PSL University, CNRS, Sorbonne Université, Paris, France \and
   Coll\`ege de France, 11 place Marcelin Berthelot, 75005 Paris
}

\date{Received 16 July 2018 / Accepted 21 May 2019}

\titlerunning{Mid-J CO lines with APEX in the northern filaments of Centaurus A}
\authorrunning{Salomé et al.}

\abstract{
   NGC 5128 (Centaurus A) is one of the best targets to study AGN-feedback in the local Universe. Optical filaments located at 16 kpc from the galaxy along the radio jet direction show recent star formation, likely triggered by the interaction of the jet with an H\rmnum{1} shell. A large reservoir of molecular gas has been discovered outside the H\rmnum{1}. In this reservoir, lies the Horseshoe complex: a filamentary structure seen in CO with ALMA and in $H\alpha$ with MUSE. The ionised gas is mostly excited by shocks, with only a minor contribution of star formation.
We used the Atacama Pathfinder EXperiment (APEX) to observe the $^{12}CO$(3-2) and $^{12}CO$(4-3) transitions, as well as dense gas tracers in the Horseshoe complex. $^{12}$CO(3-2) and $^{12}$CO(4-3) are detected for the first time in the northern filaments of Centaurus A, with integrated intensity line ratios $R_{32}\sim 0.2$ and $R_{43}\sim 0.1$, compared to the $^{12}$CO(1-0) emission. We also derived a line ratio $R_{21}\sim 0.6$, based on previous $^{12}$CO(2-1) observations.
We used the non-LTE radiative transfer code RADEX and determined that the molecular gas in this region has a temperature of $55-70\: K$ and densities between $2-6\times 10^2\: cm^{-3}$. Such densities are also in agreement with results from the Paris-Durham shock code that predicts a post-shock density of a few $100\: cm^{-3}$.
However, we need more observations of emission lines at a better angular resolution in order to place tighter constraints on our radiative models, whether they are used as a stand-alone tool (LVG codes) or combined with a shock model.}

\keywords{methods:data analysis - galaxies:individual:Centaurus A - galaxies:evolution - galaxies:ISM - galaxies:star formation - radio lines:galaxies}

\begin{document}

\maketitle


\section{Introduction}

   NGC 5128 (also known as Centaurus A) is the most nearby and well-studied radio galaxy, at a distance of 3.8 Mpc (\citealt{Harris_2010}; scaling conversion of $18.3\: pc/''$). The galaxy is surrounded by gaseous shells that have been detected in H\rmnum{1} \citep{Schiminovich_1994}, CO emission \citep{Charmandaris_2000}, and dust continuum \citep{Auld_2012}. Such gaseous shells are likely the result of a past minor merger event \citep{Malin_1983}. Along the northern radio jet, optically bright filaments are observed at distances of several kiloparsec, the so-called inner and outer filaments \citep{Blanco_1975,Graham_1981,Morganti_1991}. These filaments are thought to be the place of star formation as confirmed by Galaxy Evolution Explorer (GALEX) data \citep{Auld_2012} and young stellar clusters \citep{Rejkuba_2001}. The outer filaments are observed at the interaction of the radio jet with an H\rmnum{1} shell and were not widely studied until recently.

\begin{figure}[h]
  \centering
  \includegraphics[width=\linewidth,trim=190 35 275 85,clip=true]{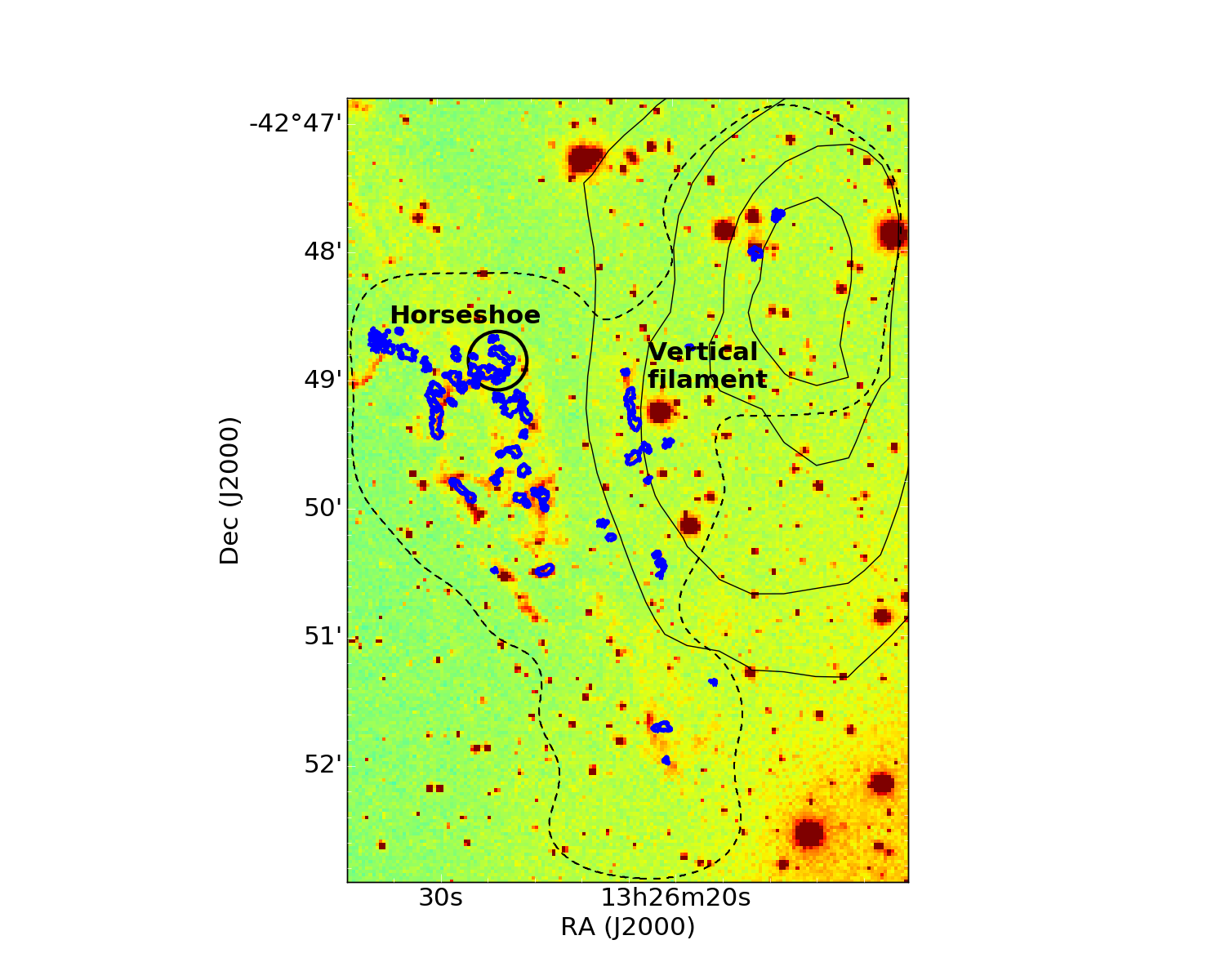}
  \caption{\label{ALMA_Ha} Distribution of the ALMA $^{12}$CO(1-0) emission without short-spacings (blue contours) in relation with the $H\alpha$ emission seen with the Cerro Tololo Inter-American Observatory (CTIO). The black contours represent the H\rmnum{1} emission and the black dashed line corresponds to the area observed with ALMA. The black circle indicates the position and size of the $27.4''$ beam of APEX for the $^{12}$CO(2-1) line, centred on the observed position.}
\end{figure}

   Recent $^{12}$CO(2-1) observations with the Atacama Pathfinder Experiment (APEX) have revealed the presence of molecular gas all along the far-ultraviolet (FUV) filaments, and the existence of a large reservoir outside the H\rmnum{1} gas \citep{SalomeQ_2016b}. Atacama Large Millimeter/submillimeter Array (ALMA) observations have shown that this large reservoir lies in a filamentary structure that we called the Horseshoe complex \citep{SalomeQ_2017}. The Horseshoe complex is associated with $H\alpha$ emission that follows the same morphology as CO (see figure \ref{ALMA_Ha}) and shares similar velocities.
Based on the Multi Unit Spectroscopic Explorer (MUSE) observations of $H\alpha$, $H\beta$, [O\rmnum{3}] and [N\rmnum{2}] in a large field encompassing the brightest regions of jet-interstellar medium (ISM) interactions, \cite{Santoro_2015b,Santoro_2016} concluded that the ionisation and kinematics of the region were the results of both fast jet entrainment and photoionisation by the UV continuum from the central engine. Combining these observations with ALMA CO and GALEX observations, and based on pixel-by-pixel BPT diagram \citep{Baldwin_1981,Kewley_2006}, \cite{SalomeQ_2016b,SalomeQ_2017} concluded that the $H\alpha$ emission detected in the Horseshoe complex originates mostly from shocks, rather than from local forming stars; there is an additional influence of the AGN in the northern part of the complex (see Fig. 8 right of Salomé et al. 2017).

   Moreover, the Horseshoe complex does not seem to be associated with recent star formation. We hypothesised that the northern filaments of Centaurus A are an example of inefficient jet-induced star formation. The jet likely enhances star formation by triggering the H\rmnum{1}-to-$H_2$ transition \citep{SalomeQ_2016b,SalomeQ_2016a} and has a $H_2$-to-H\rmnum{1} mass ratio larger than 3. However, this star formation is very inefficient, likely because of a strong injection of kinetic energy, as suggested by the relatively high virial parameter of the molecular clouds.

   We now aim to study the jet-driven energy injection in the northern filaments and its relation with the star formation efficiency. In the present paper (Paper \Rmnum{1}), we investigate the physical conditions of gas (temperature, density) and the characteristics of shocks within the Horseshoe complex. This region is where the radio jet has encountered the H\rmnum{1} shell, and has converted the atomic gas into molecular gas through compression. This is thus the ideal region to study the jet-induced star formation. To do so, we observed mid-$J_{up}$ CO transitions ($J_{up}=3,4$) and high density tracers, in the brightest CO-spot from \cite{SalomeQ_2016b}. We compare these observations with radiative transfer predictions from the RADEX non-local thermal equilibrium (LTE) radiative transfer code to constrain the temperature and local density of the gas.

   We then compare the observed ratios of the CO line integrated intensities with a small grid of the Paris-Durham shock model. We aim to show how the Paris-Durham model can be compared to extragalactic observations to infer physical conditions averaged over the rather large beam ($\sim 250\: pc$) of our single-dish observations. 
With this comparison, we also aim to understand the origin of the dynamics of the molecular gas, and its relation with the jet.

   Data from APEX are presented in section \ref{sec:Obs}. In Section \ref{sec:Res}, we analyse the data and derive line ratios of the different transitions with respect to the $^{12}$CO(1-0) from ALMA. We then compare those line ratios with predictions from RADEX and the Paris-Durham shock model in section \ref{sec:discussion}. Finally, we conclude in section \ref{sec:conclusion}.

\section{Observations}
\label{sec:Obs}

   \subsection{APEX mid-J CO}

   Submillimetre observations were made with the APEX telescope between July and December 2016 (ESO project ID 098.B-0004; PI: Salom\'e Q.). We focussed on the Horseshoe complex, and more precisely on the region with the strongest $^{12}$CO(2-1) emission discovered by \cite{SalomeQ_2016b} (so-called position 16; J2000 $\alpha=$13:26:27.47, $\delta=-$42:48:50.9). We looked for the $^{12}$CO(3-2), $^{12}$CO(4-3) and HCN/$HCO^+$(3-2) lines in one pointing, centred on the centre of position 16. We finally used the remaining observation time to look for the [C\rmnum{1}] $^3P_1$-$^3P_0$ line. The observations were made with the SHeFI/APEX-1, APEX-2, APEX-3 receivers\footnote{http://www.apex-telescope.org/heterodyne/shfi/} and backends XFFTS (bandwidths of 2.5 GHz; resolution of 88.5 kHz). Observing conditions are summarised in Table \ref{table:obs}.

\begin{table}
  \centering
  \tiny
  \begin{tabular}{lccccc}
    \hline \hline
    Line         & $\nu_{obs}$ &         FWHM          & $t_{obs}$ &  $\delta v$   & rms  \\
                 &    (GHz)    &                       &   (min)   & ($km.s^{-1}$) & (mK) \\ \hline
    CO(2-1)      &  230.1178   & $27.4''\sim 500\: pc$ &    17.8   &      3.2      &  5.6 \\
    CO(3-2)      &  345.1657   & $18.2''\sim 330\: pc$ &    23.6   &      6.4      &  6.8 \\
    CO(4-3)      &  460.2005   & $13.7''\sim 250\: pc$ &    62.0   &      6.4      &  9.2 \\
    HCN(3-2)     &  265.4018   & $23.7''\sim 430\: pc$ &   168.0   &  $\sim 11.0$  &  1.5 \\
    $HCO^+$(3-2) &  267.0699   & $23.7''\sim 430\: pc$ &   168.0   &  $\sim 11.0$  &  1.8 \\
    \CI          &  491.2637   & $12.8''\sim 230\: pc$ &    43.3   &  $\sim 12.0$  & 19.5 \\ \hline
  \end{tabular}
  \caption{\label{table:obs} Journal of observations with APEX, the $^{12}$CO(2-1) line comes from \cite{SalomeQ_2016b}. The rms were determined with both polarisations and are given in main beam temperature for the indicated spectral resolution.}
\end{table}

   The data were reduced using the IRAM package CLASS. After dropping bad spectra, a linear baseline was subtracted from the average spectrum, except for the $^{12}$CO(4-3) average spectrum for which we substracted a degree 4 polynomial to correct for baseline oscillations. For detections, the baseline was subtracted at velocities outside the range of the emission line. Each spectrum was smoothed to a spectral resolution of $\sim 3-12\: km.s^{-1}$ (Table \ref{table:obs}).

   \subsection{ALMA $^{12}$CO(1-0) short-spacings with ACA}

   In \cite{SalomeQ_2017}, we presented $^{12}$CO(1-0) observations with the ALMA 12m array. The data were taken during Cycle 3 using Band 3 receivers. We mapped a region of $6.1'\times 4.3'$ with a mosaic of 34 pointings (integration time between 140 and 430s), at a resolution of $1.30''\times 0.99''\sim 23.8\times 18.1\: pc$ ($PA=81.5$\degree).

   The maximum recovered angular scale by the 12m array was $14''\sim 260\: pc$. To recover the short-spacings, we also observed the northern filaments with the Atacama Compact Array 7m array (ACA) using Band 3 receivers during Cycle 3 (project ADS/JAO.ALMA$\#$2015.1.01019.S; PI: Salom\'e Q.). The map consists in a mosaic of 15 pointings, each with an integration time between 11.7 and 41 min, covering the same region as the ALMA observations. The baselines of the 7m array range from 8.85m to 48.95m, which produces a synthesised beam of $13.68''\times 8.28''\sim 250.3\times 151.5\: pc$ ($PA=90.5$\degree).

\begin{figure*}[h]
  \centering
  \includegraphics[height=9.5cm,trim=190 35 275 85,clip=true]{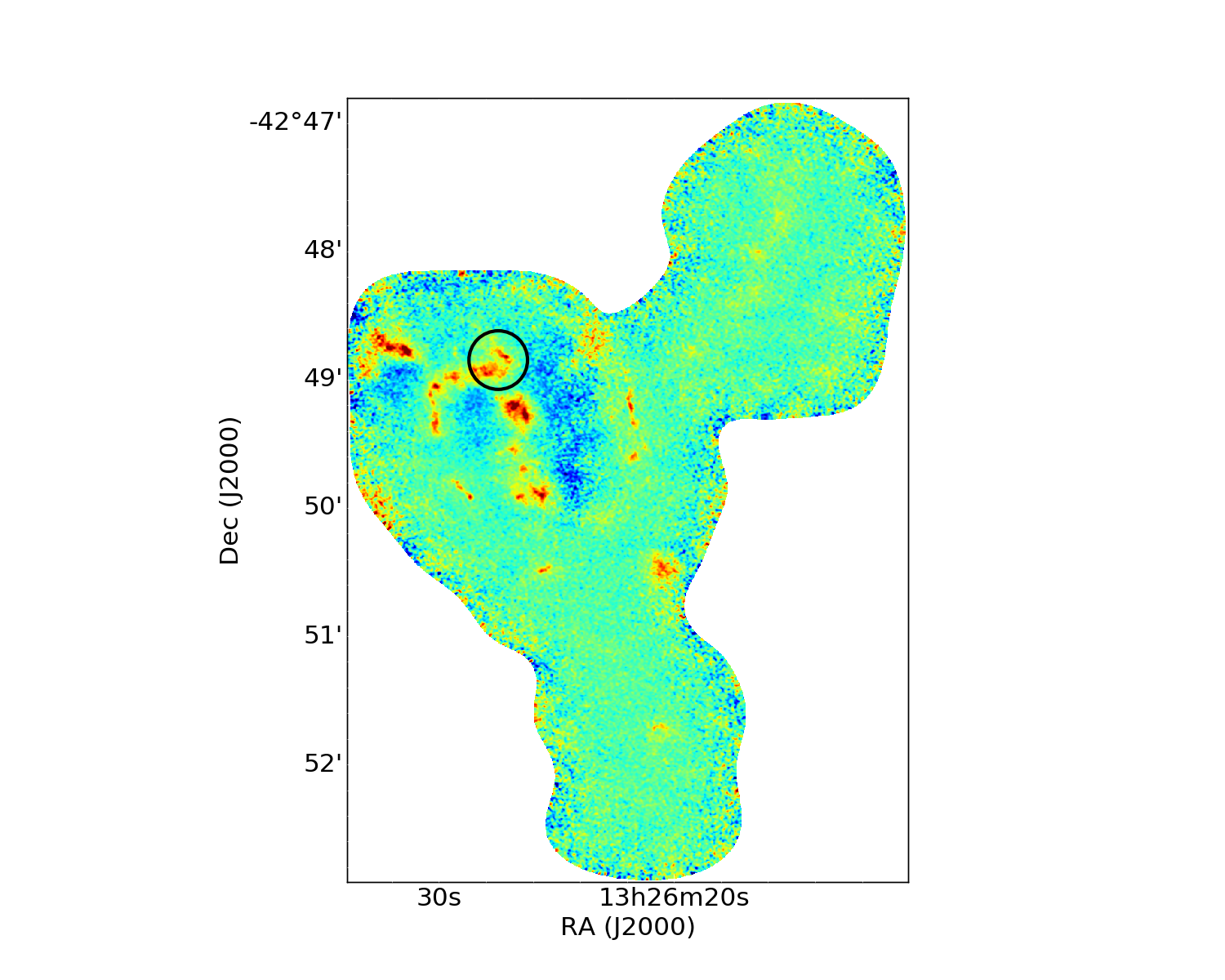}
  \hspace{3mm}
  \includegraphics[height=9.5cm,trim=290 35 190 85,clip=true]{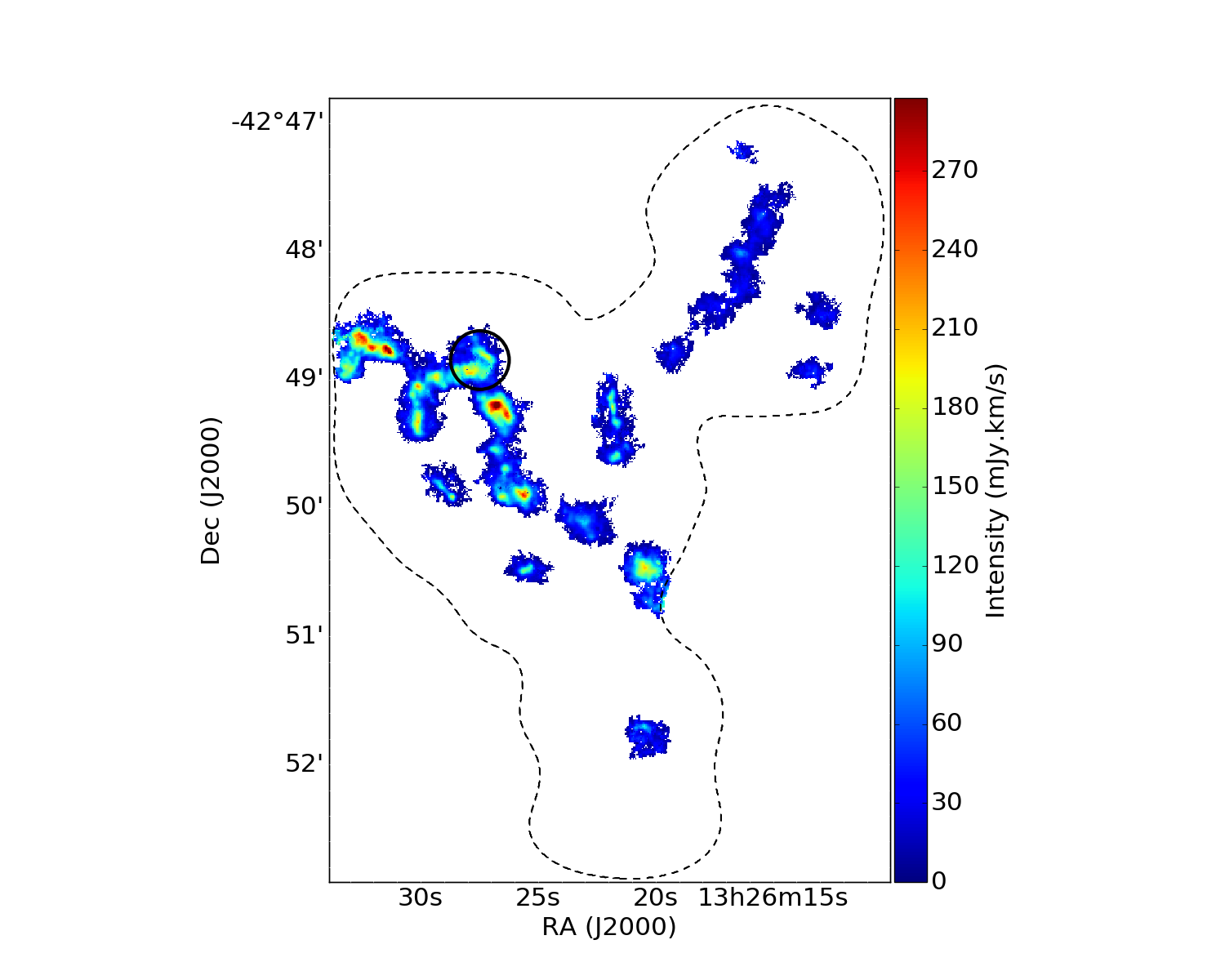}
  \caption{\label{moment0_ALMA} \emph{Left:} Map of the moment 0 in $Jy/beam.km.s^{-1}$ of the ALMA+ACA data produced with CASA. \emph{Right:} Integrated $^{12}$CO intensity map in the velocity range $-350<v<-100\: km.s^{-1}$. The dashed contours represent the region observed with ALMA+ACA. In both images, the black circle indicates the target position of this paper ($27.4''$ beam of APEX).}
\end{figure*}

   Both sets of data were calibrated using the Common Astronomy Software Applications (CASA) pipeline. We then combined the data in the uv plane within CASA\footnote{https://casaguides.nrao.edu/index.php/M100\_Band3}. The imaging and cleaning was made using the \emph{tclean} routine in CASA. The resolution is slightly smaller than that of the ALMA 12m data alone, with a synthesised beam of $1.40''\times 1.10''\sim 25.6\times 20.1\: pc$ ($PA=82.9$\degree). Figure \ref{moment0_ALMA} shows the moment 0 map of the combined data. The histogram of the noise level peaks at 6.2 mJy/beam at a spectral resolution of $1.47\: km.s^{-1}$.

\section{Results}
\label{sec:Res}

   \subsection{APEX observations}

   $^{12}$CO(3-2) emission is well detected at $>5\sigma$ with APEX and the spectrum reveals two velocity components around $-230$ and $-270\: km.s^{-1}$, relative to Centaurus A, with widths of about $40\: km.s^{-1}$. Similarly, the $^{12}$CO(2-1) emission was re-reduced and decomposed in two velocity components (figure \ref{spectra}). The redder component is stronger than the bluer component. We also detected $^{12}$CO(4-3) emission at $>3\sigma$, but the presence of two velocity components is less clear.
We integrated the intensities over two ranges of velocity: $-360<v<-260\: km.s^{-1}$ for the blue component and $-260<v<-160\: km.s^{-1}$ for the red component. These two velocity ranges were determined by fitting the CO(2-1) line profile with two Gaussians. We summarise the characteristics of the $^{12}$CO emission in Table \ref{table:spec}. In the following, the intensities used to derive the line ratios are always expressed in $K.km.s^{-1}$.

\begin{figure*}[h]
  \centering
  \includegraphics[page=2,height=8cm,trim=35 30 85 85,clip=true]{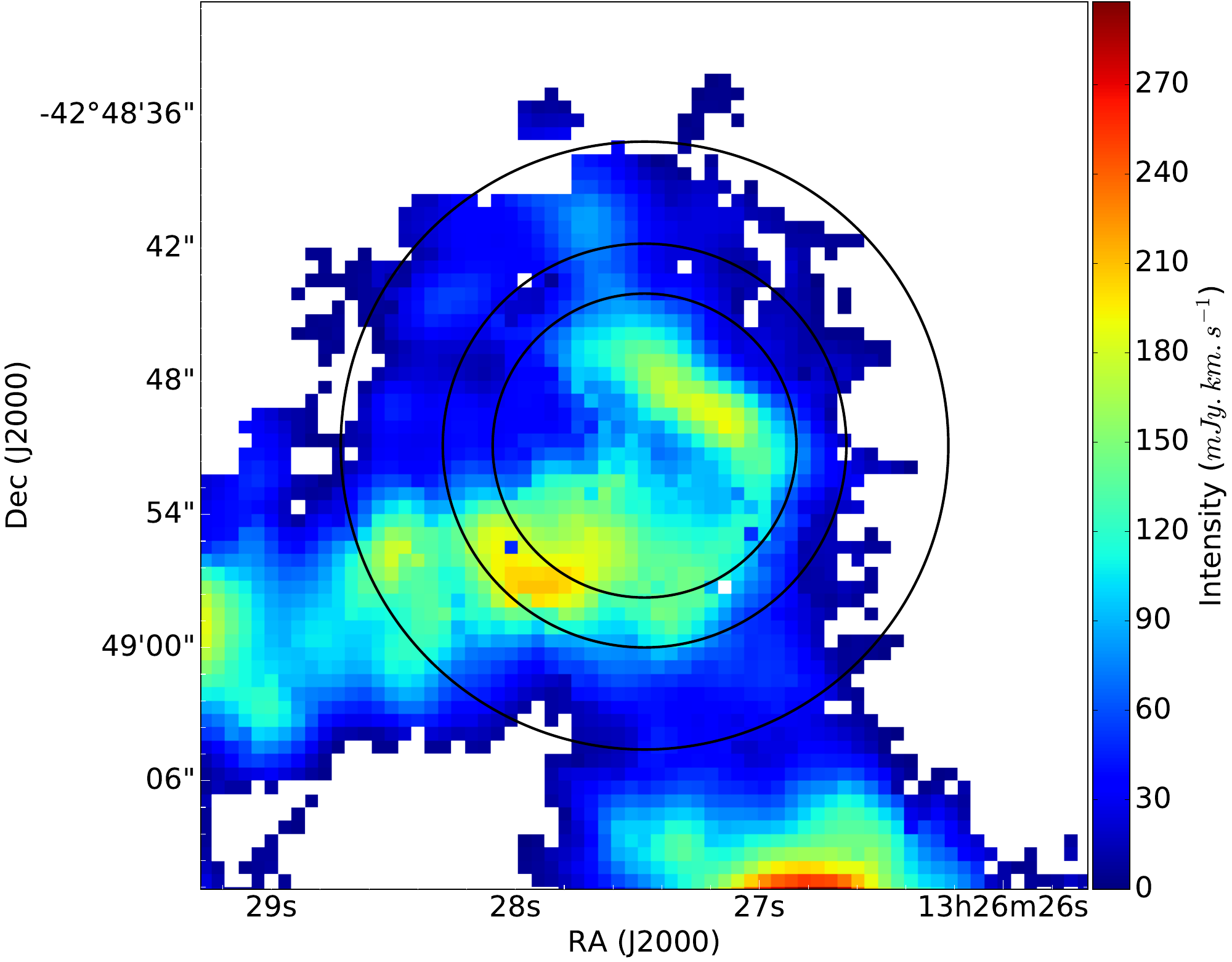}
  \hspace{5mm}
  \includegraphics[page=3,height=8cm,trim=205 30 180 85,clip=true]{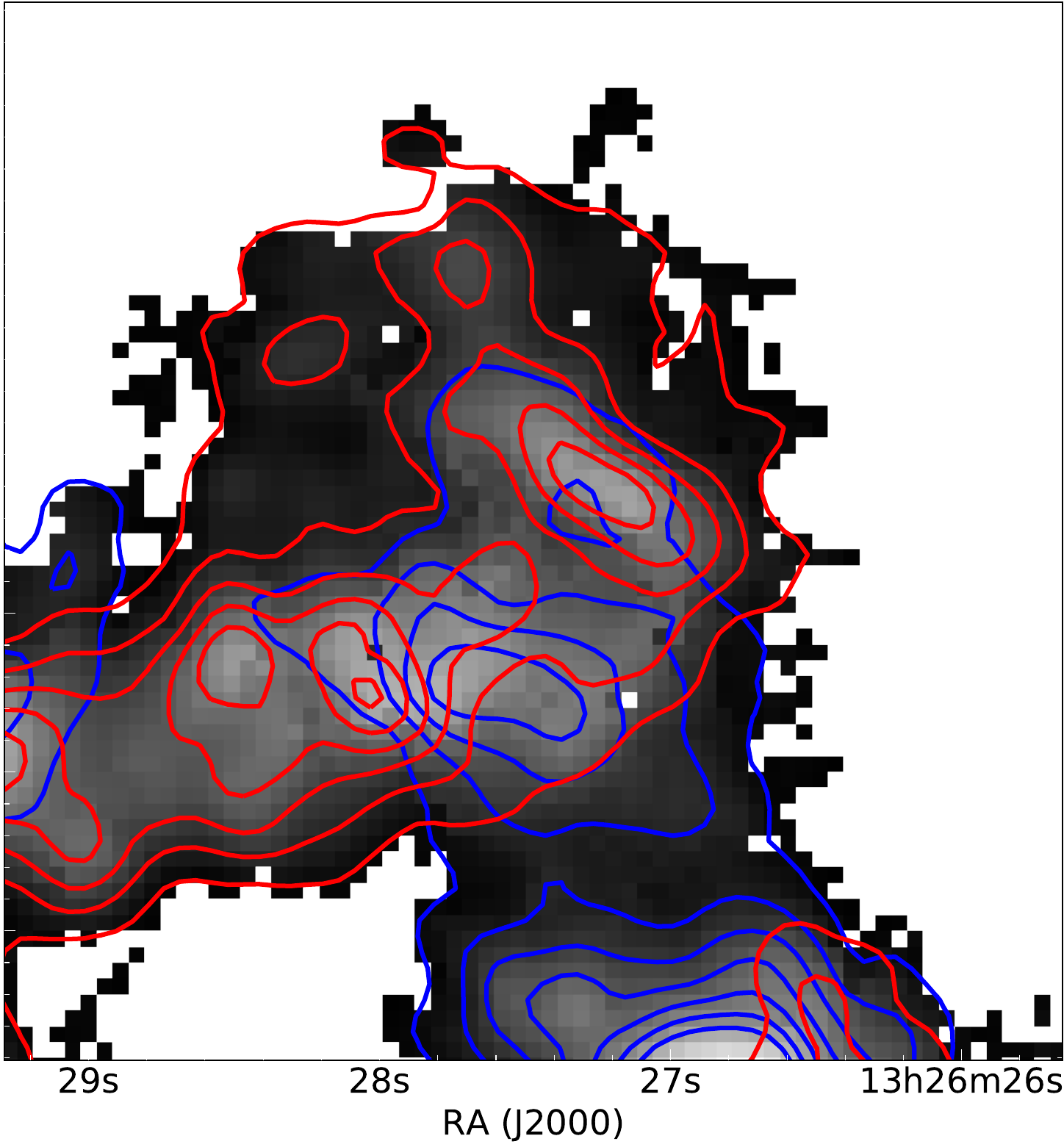}
  \caption{\label{velo-clouds} \emph{Left:} Intensity map of the $^{12}$CO(1-0) emission as seen by ALMA+ACA in the region observed with APEX. From outer to inner, the circles correspond to the beam of APEX for the $^{12}$CO(2-1), (3-2) and (4-3). \emph{Right:} Distribution of the two ALMA velocity components. Contours of each component intensity are overlaid on the total intensity map (grey colour).}
\end{figure*}

\begin{table}[h]
  \centering
  \small
  \begin{tabular}{lccccc}
    \hline \hline
    Line              & $\int T_{mb}d\varv$ (line) & $\int T_{mb}d\varv$ (CO10) &   $R_{line}$   \\
                      &      ($K.km.s^{-1}$)       &      ($K.km.s^{-1}$)       &                \\ \hline
    CO(2-1) blue      &       $3.18\pm 0.49$       &       $4.57\pm 0.69$       & $0.70\pm 0.21$ \\
    CO(2-1) red       &       $4.43\pm 0.67$       &       $7.91\pm 1.19$       & $0.56\pm 0.17$ \\
    Total             &       $7.61\pm 1.16$       &      $12.48\pm 1.88$       & $0.61\pm 0.18$ \\ \hline
    CO(3-2) blue      &       $1.36\pm 0.26$       &       $5.57\pm 0.84$       & $0.24\pm 0.08$ \\
    CO(3-2) red       &       $2.29\pm 0.38$       &      $12.22\pm 1.83$       & $0.19\pm 0.06$ \\
    Total             &       $3.65\pm 0.64$       &      $17.79\pm 2.67$       & $0.21\pm 0.07$ \\ \hline
    CO(4-3) blue      &       $0.51\pm 0.24$       &       $6.77\pm 1.02$       & $0.08\pm 0.05$ \\
    CO(4-3) red       &       $1.68\pm 0.34$       &      $15.69\pm 2.35$       & $0.11\pm 0.04$ \\
    Total             &       $2.19\pm 0.58$       &      $22.46\pm 3.37$       & $0.10\pm 0.04$ \\ \hline
    HCN(3-2) blue     &          $<0.24$           &       $4.84\pm 0.73$       &    $<0.05$     \\
    HCN(3-2) red      &          $<0.24$           &       $9.28\pm 1.39$       &    $<0.03$     \\
    Total             &          $<0.48$           &      $14.12\pm 2.12$       &    $<0.03$     \\ \hline
    $HCO^+$(3-2) blue &          $<0.29$           &       $4.84\pm 0.73$       &    $<0.06$     \\
    $HCO^+$(3-2) red  &          $<0.29$           &       $9.28\pm 1.39$       &    $<0.03$     \\
    Total             &          $<0.58$           &      $14.12\pm 2.12$       &    $<0.04$     \\ \hline
    \CI\, blue        &          $<3.10$           &       $7.07\pm 1.06$       &    $<0.44$     \\
    \CI\, red         &          $<3.10$           &      $16.50\pm 2.48$       &    $<0.19$     \\
    Total             &          $<6.20$           &      $23.57\pm 3.54$       &    $<0.26$     \\ \hline
  \end{tabular}
  \caption{\label{table:spec} Integrated intensity of the lines observed with APEX, and the $^{12}$CO(1-0) emission observed with ALMA+ACA, at the corresponding APEX resolution.
The intensities were derived by adding all the channels in the velocity ranges $-360<v<-260\: km.s^{-1}$ (blue) and $-260<v<-160\: km.s^{-1}$ (red), relative to Centaurus A. For non-detections, an upper limit at $3\sigma$ has been derived assuming a linewidth of $50\: km.s^{-1}$, that is the typical linewidth of the $^{12}$CO(1-0) line. The last column gives the ratio of the line with the $^{12}$CO(1-0).}
\end{table}

   Contrary to the CO lines, we did not detect the HCN and $HCO^+$(3-2) lines at the present noise level. The [C\rmnum{1}] line was not detected either. However, this is due to the poor noise level we reached in the small amount of time we observed this line . For those three lines, we estimated upper limits at $3\sigma$, assuming a line width of $50\: km.s^{-1}$.

   \subsection{Line ratios}

   To study the excitation of molecular gas and its properties (temperature, density), we used the ratio between the different transitions. However, we cannot compare our APEX observations between themselves because of the different beam sizes (and since we only observed one position). It is therefore impossible to accurately estimate the line ratios with our APEX observations alone. To overcome this difficulty, we used our high-resolution $^{12}$CO(1-0) observations obtained with ALMA+ACA to determine the $^{12}$CO(1-0) emission contained in the different APEX beams (see figure \ref{velo-clouds}).

   We first recovered the signal using a coherent Gaussian decomposition method, based on the method of \cite{MAMD_2017a}. We then used the clustering method of \citeauthor{MAMD_2017a}, which is based on a threshold descent applied to a cube of integrated flux. Because of the low signal-to-noise ratio, the signal is difficult to recover completely. Therefore, we first decreased the spatial resolution to $2.2''$.
Once we extracted the signal from the data, we reconstructed a cube of the modelled $^{12}$CO(1-0) emission. We then applied the ALMA K/Jy conversion factor of 19.17 to obtain a cube of the main beam temperature.
To get the $^{12}$CO(1-0) emission contained in the APEX beam, we smoothed the reconstructed $^{12}$CO(1-0) cube to the different APEX resolutions and extracted the spectrum at the central position of the APEX observations. At every resolution, the $^{12}$CO(1-0) spectrum shows two distinct velocity components. As for the APEX observations, we integrated the spectra in the same velocity ranges to estimate the integrated intensity of both components (see Table \ref{table:spec}). Figure \ref{spectra} enables a quick comparison of the profiles obtained with APEX and the $^{12}$CO(1-0) at the same resolution.
\bigskip

   We were thus able to estimate the integrated intensity $I\, [K.km.s^{-1}]$ line ratio of the different transitions observed with APEX compared to the $^{12}$CO(1-0) emission. The line ratios are reported in Table \ref{table:spec}, i.e.
\begin{equation}
  R_{line}=I_{line}/I_{\rm CO10}
\end{equation}

   For the different CO lines, we estimated two intensity line ratios, one for each velocity component. For the blueshifted component, we found line ratios of about 0.7, 0.25, and 0.1 for the $^{12}$CO(2-1), $^{12}$CO(3-2), and $^{12}$CO(4-3), respectively. The red component presents lower lines ratios of about 0.55, 0.2, and 0.1. The line ratios are consistant within the error bars. However, such difference in the line ratios may be related to different excitation between the two velocity components.

   The typical CO(2-1)/CO(1-0) line ratio in the Milky Way and nearby galaxies spans between $0.6-1.0$ with a typical value of 0.8 \citep{Hasegawa_1997, Sakamoto_1997,Oka_1998,Leroy_2009,Vlahakis_2013}. The observed CO(2-1)/CO(1-0) line ratio is consistent with those values. We also found CO(3-2)/CO(1-0) and CO(4-3)/CO(1-0) line ratios consistent with the ratios found in the Milky Way and M51 \citep{Nieten_1999,Kim_2002,Carilli_2013, Vlahakis_2013}.

   For the other lines, we used the $3\sigma$ upper limits to derive upper limits of the corresponding line ratios. For each line, we derived two upper limits, one for each velocity component. The line ratios of the dense gas tracers have upper limits of about $3-6\%$ when compared with the $^{12}$CO(1-0) emission. Such non-detection may be explained by the fact that the dense molecular gas is more compact and the emission is diluted by the beam. Another possible explanation is that very little HCN/$HCO^+$(3-2) emission is produced in this region because of an insufficient gas density to excite these transitions ($n_{crit}\sim 10^6-10^7\: cm^{-3}$; see below).

   Finally, the non-detection of the [C\rmnum{1}] line is not a surprise. Indeed, we derived upper limits of the [C\rmnum{1}]/CO(1-0) line ratio of $20-35\%$. This is higher than the typical [C\rmnum{1}]/CO(1-0) line ratio of 20\% found in galaxies at low and high redshift \citep{Gerin_2000,Weiss_2005, Jiao_2017}. Therefore, deeper observations are clearly necessary for the atomic carbon detection.

\section{Discussion}
\label{sec:discussion}

   \subsection{Low-velocity gradient modelling of the CO emission}

\begin{figure*}
  \centering
  \includegraphics[page=5,height=6cm,trim=12 16 33 30,clip=true]{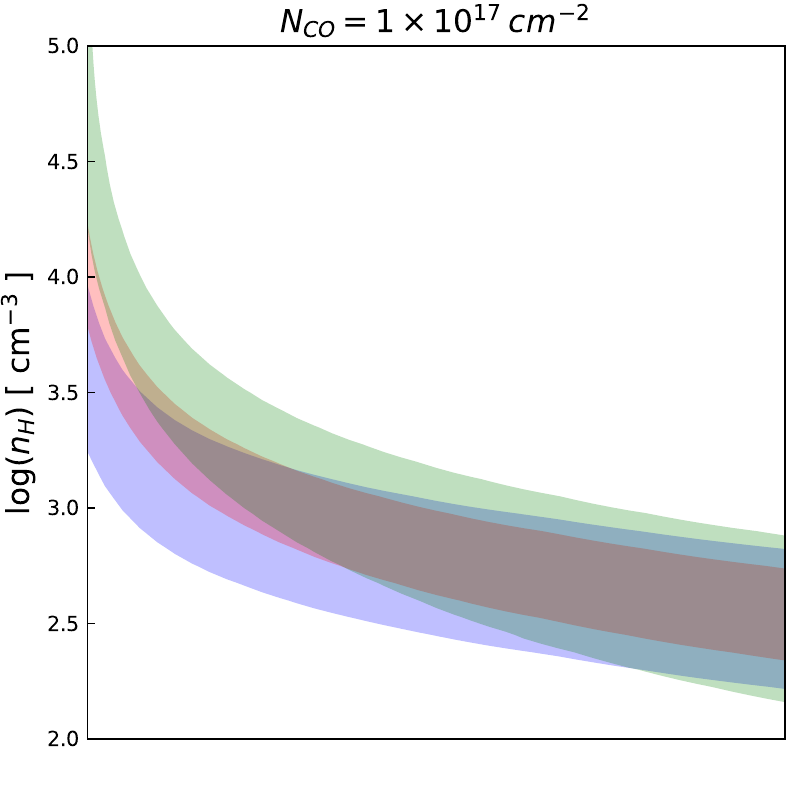}
  \hspace{-2mm}
  \includegraphics[page=5,height=6cm,trim=12 16 33 30,clip=true]{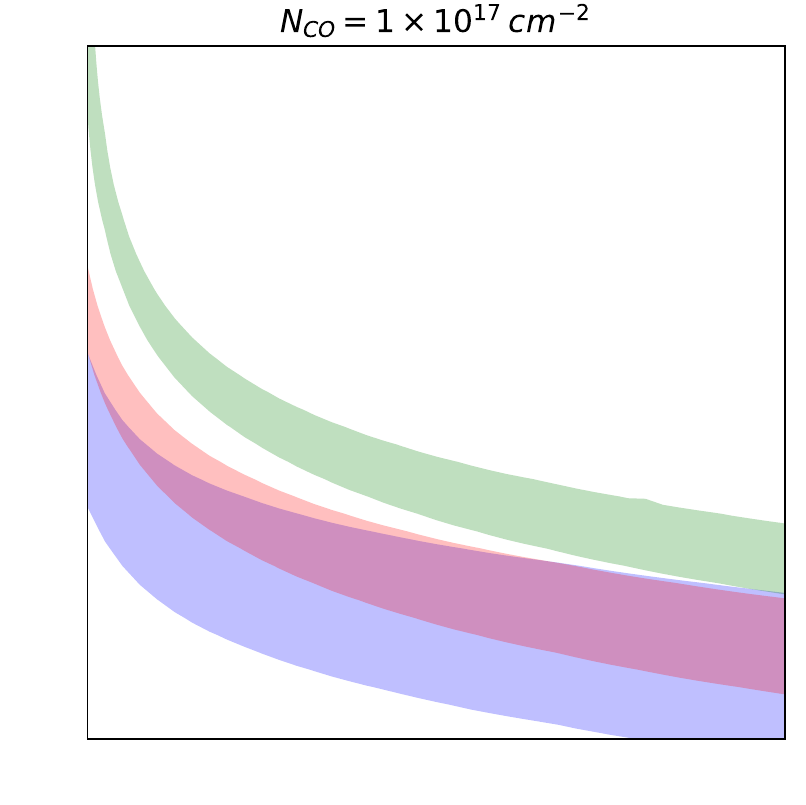}
  \hspace{-2mm}
  \includegraphics[page=5,height=6cm,trim=12 16 33 30,clip=true]{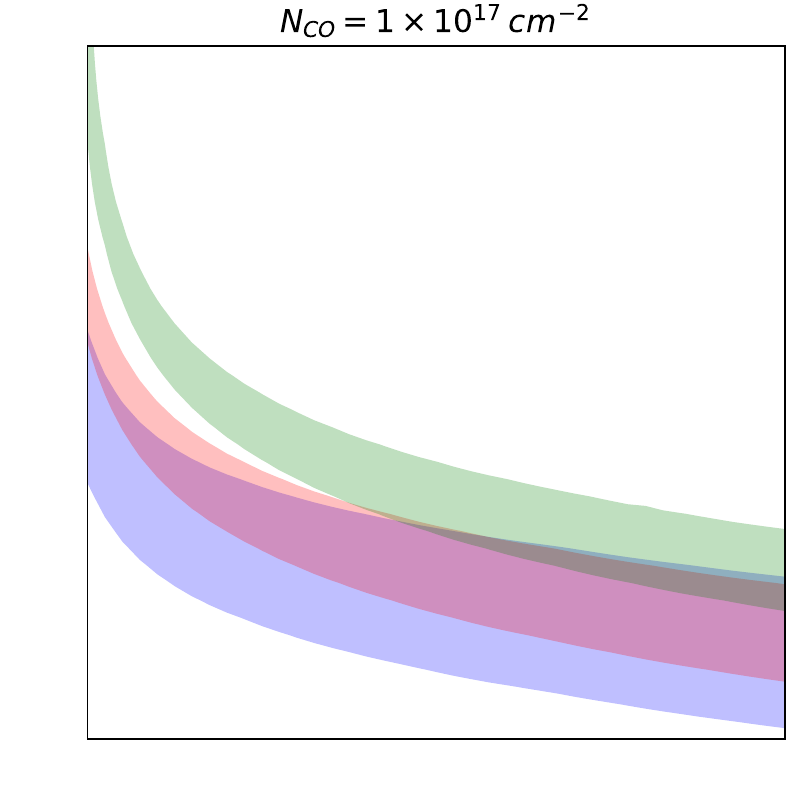} \\
  \vspace{-3mm}
  \includegraphics[page=6,height=6cm,trim=12 16 33 30,clip=true]{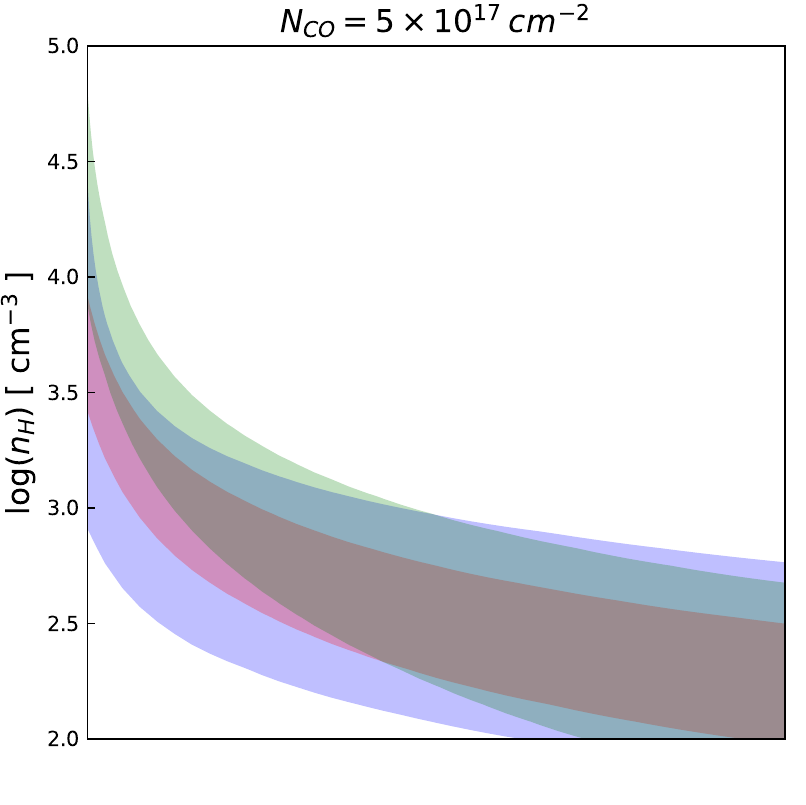}
  \hspace{-2mm}
  \includegraphics[page=6,height=6cm,trim=12 16 33 30,clip=true]{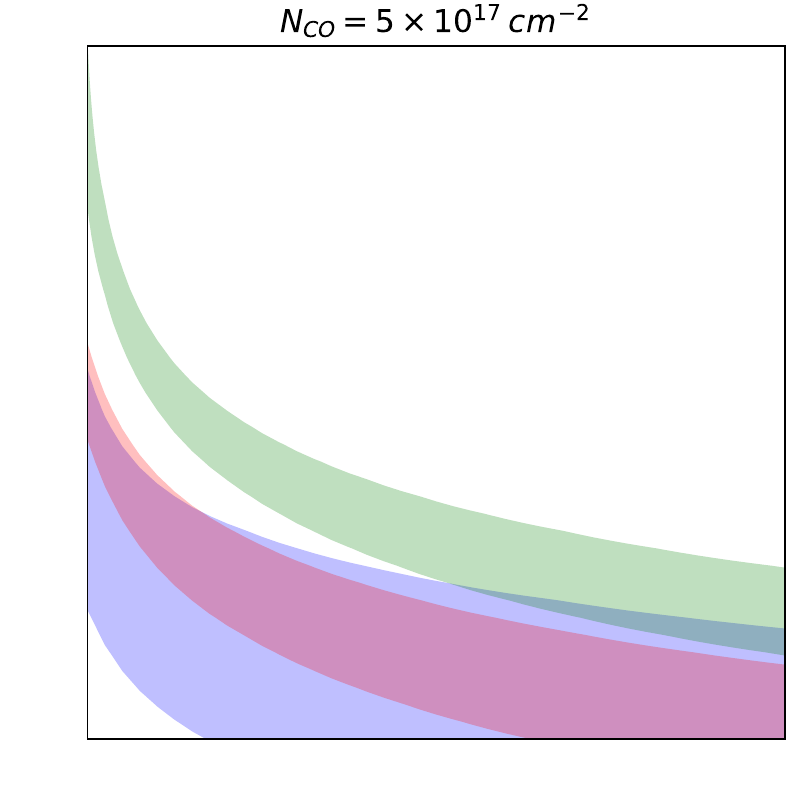}
  \hspace{-2mm}
  \includegraphics[page=6,height=6cm,trim=12 16 33 30,clip=true]{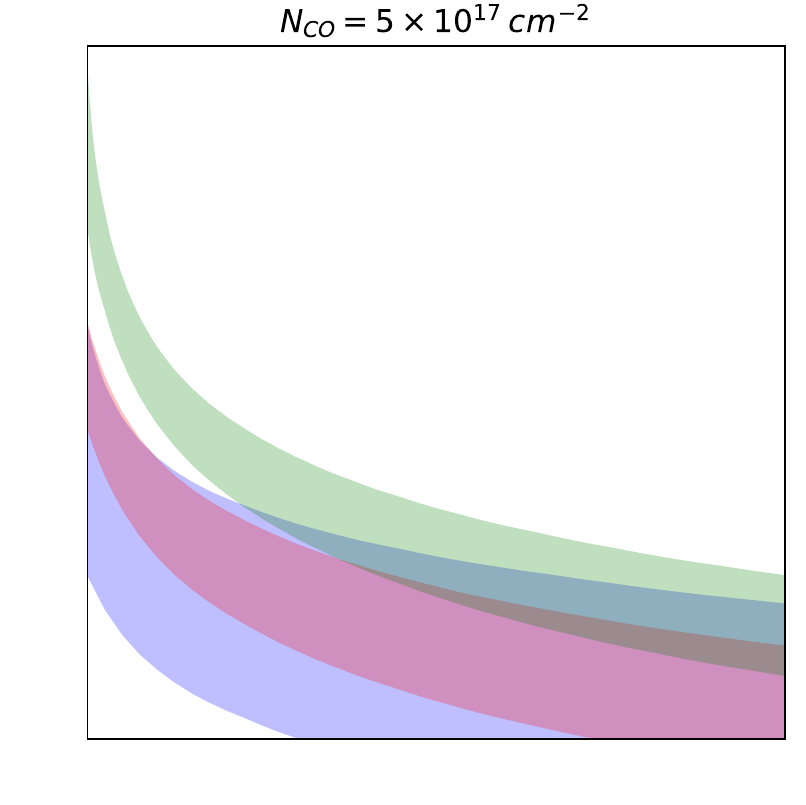} \\
  \vspace{-3mm}
  \includegraphics[page=7,height=6cm,trim=12 16 33 30,clip=true]{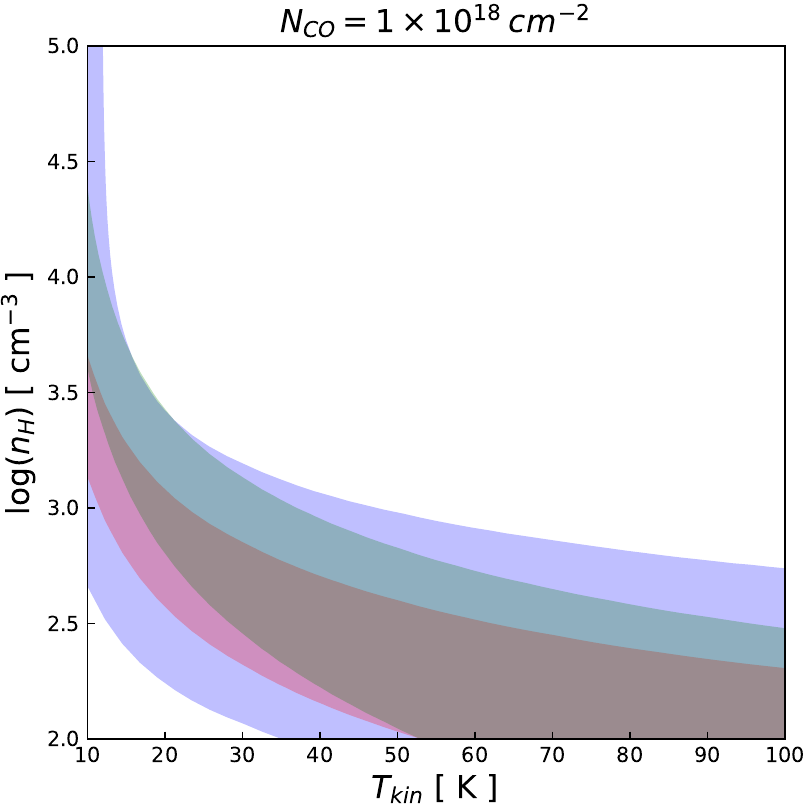}
  \hspace{-2mm}
  \includegraphics[page=7,height=6cm,trim=12 16 33 30,clip=true]{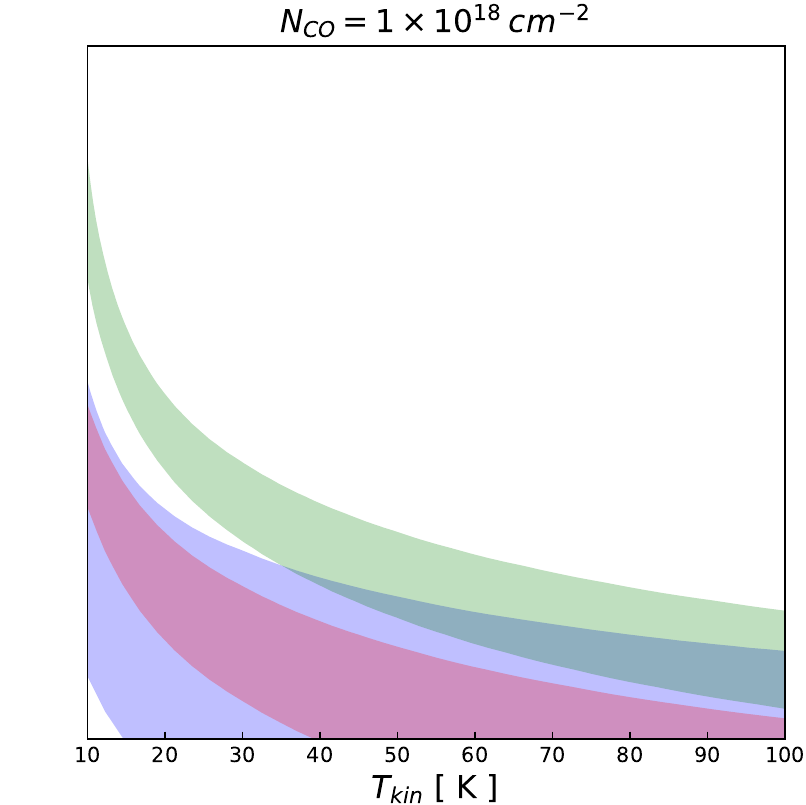}
  \hspace{-2mm}
  \includegraphics[page=7,height=6cm,trim=12 16 33 30,clip=true]{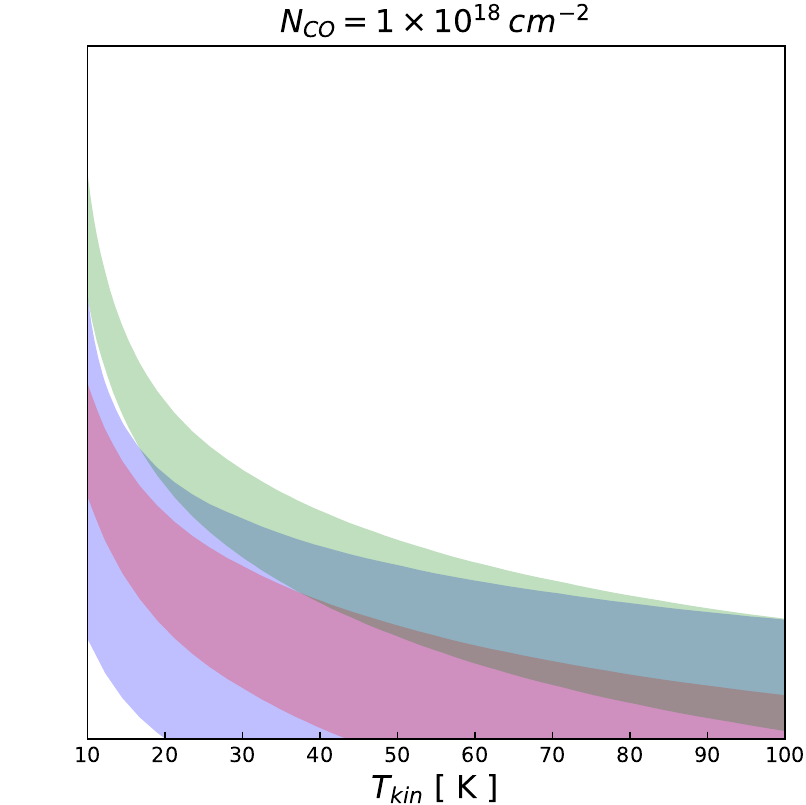} \\
  \caption{\label{params} Radiative transfer predictions of the CO intensity line ratios as a function of the temperature and density. The blue, red, and green areas show the observed CO(2-1)/CO(1-0), CO(3-2)/CO(1-0), and CO(4-3)/CO(1-0) ratio. The darkest area corresponds to the models which fit the observations.
  From \emph{left} to \emph{right}, the columns correspond to the blueshifted velocity component, the redshifted velocity component and all the emission, respectively.}
\end{figure*}

   We used the non-LTE radiative transfer code RADEX \citep{RADEX} based on the low-velocity gradient (LVG) approximation. For each triplet \{$n_H$, $N_{CO}$, $T_{kin}$\}, we built a CO-SLED model ($40\times 8\times 40$). We then only kept the \{$n_H$, $N_{CO}$, $T_{kin}$\} that are consistent with the observed line ratios within their error bars. We thus restricted the parameter space for (i) $n_H$ from $10^2-10^5\: cm^{-3}$ to $n_H=200-1000\: cm^{-3}$ (ii) for $N_{CO}$ from $10^{15}-10^{19}\: cm^{-2}$ to $10^{17}-10^{18}\: cm^{-2}$ and (iii) for T from $10-100\: K$ to $25-70 K$. We fixed a linewidth $\Delta V=50\: km.s^{-1}$, which gives $N_{CO}/\Delta V=2\times 10^{15}-2\times 10^{16}\: cm^{-2}.(km.s^{-1})^{-1}$.

   Now that the parameter space has been reduced by excluding the incompatible triplets of parameters \{$n_H$, $N_{CO}$, $T_{kin}$\}, we searched for the best among the possible triplets. Since the models are poorly constrained, we coarsely sampled the kinetic temperatures between 25 and 70 K by step of 5 K, CO column densities $N_{CO}=10^{17},5\times 10^{17},10^{18}\: cm^{-2}$ and total volume densities $n_H=2,4,6,8,10\times 10^2\: cm^{-3}$.


\begin{figure}[h]
  \centering
  \includegraphics[height=7cm,trim=40 10 50 35,clip=true]{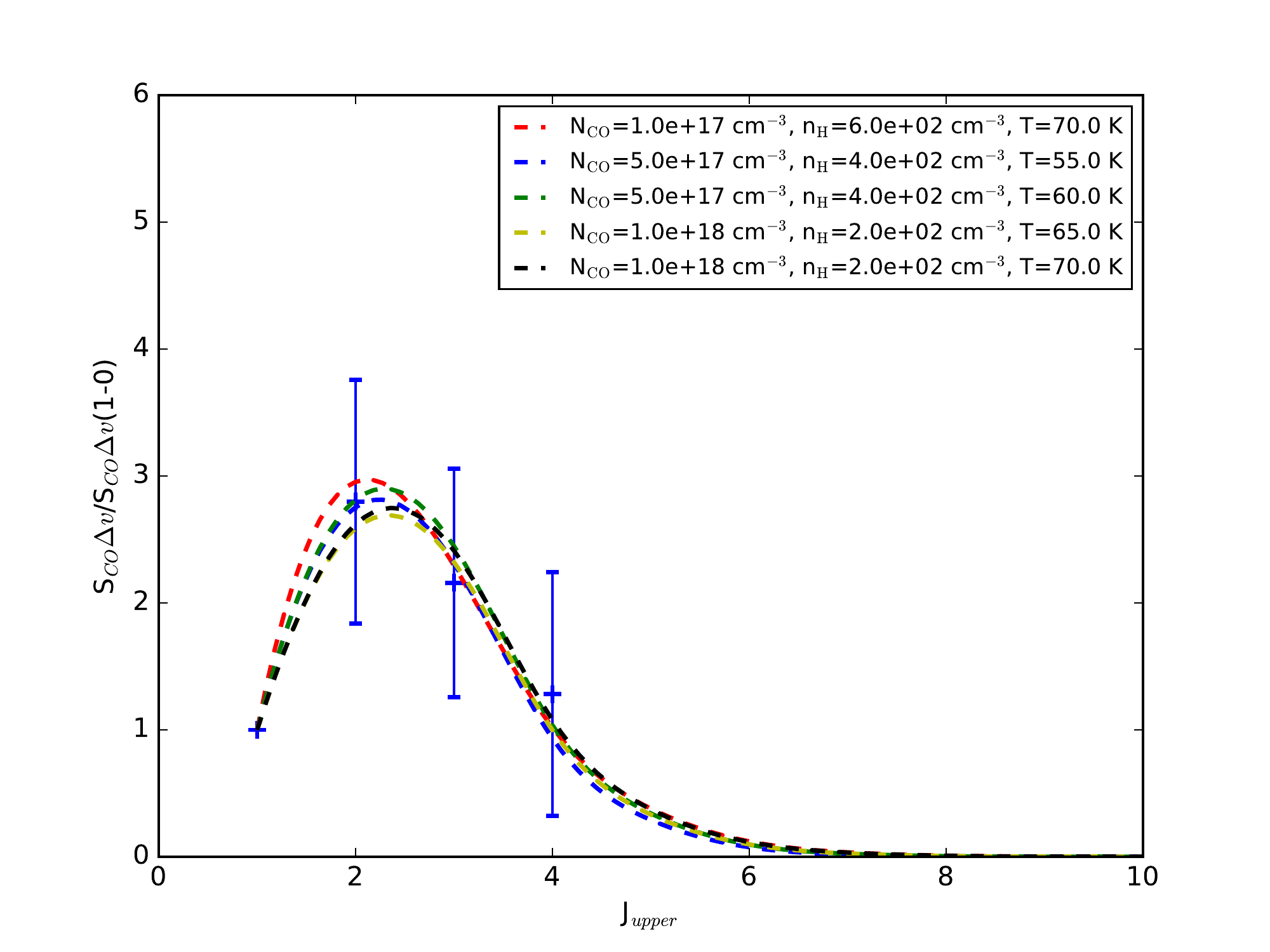}
  \caption{\label{SLED} Spectral line energy distribution of the blueshifted component. The coloured dashed lines represent the best model predictions for different triplets \{$n_H$, $N_{CO}$, $T_{kin}$\}.}
\end{figure}

   We assume that the excitation of the different clouds is the same in this region. We consider a model as reliable if the predicted line ratios are within the error bars. Plotting the parameter space (figure \ref{params}), we see that the solution is not unique. We thus computed a $\chi^2$ between those model predictions and the observed line ratios. We identified a range of solutions that minimise this $\chi^2$, which are for $N_{CO}=10^{17}-10^{18}\: cm^{-2}$, $n_H=2-6\times 10^2\: cm^{-3}$ and $T_{kin}=55-70\: K$ (figure \ref{SLED}). This are the best constrains that we could derive with the current data. For a given $N_{CO}$, the higher the volume density, the lower the temperature. Moreover, it appears that, for a given temperature, the blueshifted component is denser than the redshifted component. Similarly, the blueshifted component tends to be warmer for a given density.

   The present study is the best we can do with the current data. Nevertheless, we are aware of number of caveats that we discuss in the following. We know that the emission contained in the beam of APEX comes from a collection of molecular clouds. Therefore, RADEX only constrains the average density of these clouds. But it is very likely that the density range of the molecular clouds spans over a larger range when considered individually, and the density predicted by RADEX is affected by the filling factor. Moreover, the fraction of diffuse-to-dense gas within the APEX beam needs to be taken into account. Finally, as we have only one pointing for each transition and the beam size is not the same, we do not cover the same regions. It is thus possible that we introduce a bias in the line ratios.
All these factors can explain why we found a degeneracy between $(n,\: N,\: T)$. More observations are needed to lift this degeneracy, in particular high-resolution observations with ALMA+ACA for several CO transitions. Indeed, only this will allow us to take only the regions emitting in the different lines into account (see Paper \Rmnum{2} for a preliminary study).

   We compared the observed HCN/$HCO^+$(3-2) and [C\rmnum{1}] upper limits with the predictions from RADEX. We ran the same grid as for the CO lines, but with different column densities. For HCN and $HCO^+$, we used the relative abundances $HCN/CO\sim 10^{-4}$ and $HCO^+/CO\sim 3\times 10^{-5}$ from \cite{Pety_2017}. For the [C\rmnum{1}], we assumed a column density of 0.1 times the CO column density, similar to what is commonly observed in nearby galaxies \citep{Gerin_2000,Sofia_2004,Bolatto_2013}. However, these upper limits do not provide additional constraints on the physical conditions of gas.

   \subsection{Paris-Durham shock code}

   We then attempted to interpret our observations by combining a radiative transfer treatment based on the LVG approximation with a shock model. The shock model is the Paris-Durham code \citep{Flower_2015}. This simulates the propagation of a one-dimensional shock wave through the ISM and self-consistently calculates the physical, dynamical and chemical structure of a shocked layer. The radiative transfer code dealing with CO lines has been presented in \cite{Gusdorf_2008}, where it was used to perform comparisons with observations of shocked CO in protostellar outflows. The application of such a combination of models to observations was later systematised in the study of shocks in the W28F supernova remnant \citep{Gusdorf_2012}. In such a Galactic environment, this method yields accurate constraints on physical parameters, and enables the subsequent characterisation of shock energetics (momentum and energy injected by the shocks in interacting clouds). Our aim is to better understand the nature of the shocks that could propagate and emit bright CO lines.

   We attempted to compare the observed ratios of CO line integrated intensities with the results of a small grid of models. This kind of grid was first used in an extragalactic environment in \cite{Lee_2016b} in the N159W region of the Large Magellanic Cloud. It consists of an ensemble of stationary C-type and J-type shock models with pre-shock densities $n_H=10^2,\: 5\times 10^2,\: 10^3,\: 5\times 10^3,\: 10^4,\: 5\times 10^4,\: 10^5,\: 5\times 10^5,\: 10^6\: cm^{-3}$. The shock velocities ($\varv_s$) were varied between 4 and $20\: km.s^{-1}$ for C-type models and between 4 and $30\: km.s^{-1}$ for J-type models; by definition, the velocity of a C-type shock is limited by the magnetosonic velocity of the medium where it propagates. The external radiation field characterised by a scaling factor '$G_0$' was set equal to 0 or 1 with respect to the mean interstellar radiation field as described by \cite{Draine_1978}. As a first step, our aim is to start placing constraints on these input parameters of the model thanks to the comparison with observations. After such an exploration and with more observations, we will fine-tune other input parameters. This will be the case of, for example, the magnetic field strength perpendicular to the shock propagation. In the current grid, this quantity was set as $B_\bot\, (\mu G)=[n_H\, (cm^{-3})]^{1/2}$.

\begin{figure*}[h]
  \centering
  \includegraphics[width=\linewidth]{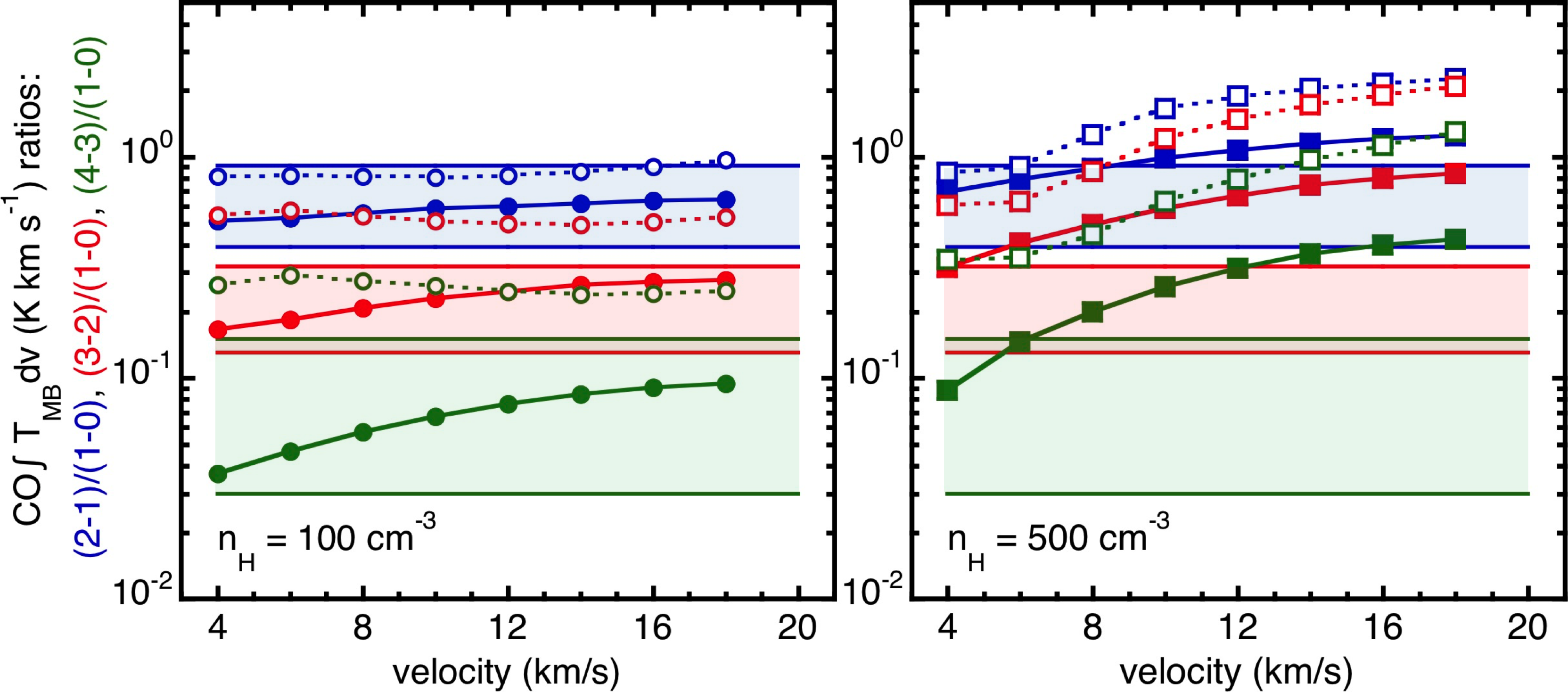}
  \caption{\label{shock_models} Results obtained with the Paris-Durham shock model. Each panel corresponds to a different pre-shock density: $n_H=100$ (\emph{left}) and $500\: cm^{-3}$ (\emph{right}). The three coloured rectangles (blue, red, and green) show the observational results: the (2-1)/(1-0), (3-2)/(1-0), and (4-3)/(1-0) ratios, respectively (from Table \ref{table:spec}). Each point corresponds to a set of \{$n_H$, $G_0$, and $\varv_s$\} values; $\varv_s$ values are indicated in the X-axis. The model results are shown with symbols: filled or empty for $G_0=0$ or 1, respectively.}
\end{figure*}

   We present our preliminary results in Figure \ref{shock_models}. In this figure, we compare the observations of the total line integrated intensity ratios ('Total' lines in Table \ref{table:spec}) with the results of models. We only present results obtained with C-type models with $n_H=100$ (left panel) and $500\: cm^{-3}$ (right panel). The three blue, red, and green rectangles show the observational results: they correspond to the area between the minimum and maximum $R_{line}$ values from Table \ref{table:spec} for the (2-1)/(1-0), (3-2)/(1-0), and (4-3)/(1-0) ratios, respectively. The model results are shown with symbols: circles and squares for $n_H=100$ and $500\: cm^{-3}$, and filled or empty for $G_0=0$ or 1. Each point corresponds to a set of \{$n_H$, $G_0$, and $\varv_s$\} values, where $\varv_s$ values are indicated in the X-axis. The figure shows the principle, successes, and current shortcomings of our modelling. A significant part of our grid does not yield satisfying fits to the observed ratios, so that we only show the best-fitting models. For instance, shock models with higher pre-shock densities yield too high ratios, and J-type models produce absolute integrated intensities that are too low compared with observed values; J-type shocks are not as wide as C-type shocks and generate CO column densities that are too low. Among the models that are shown, the best models are obtained for $n_H=100\: cm^{-3}$ and $G_0=0$. This result means that, on average, the total gas encompassed by the beam behaves like the gas processed by a shock with these characteristics. In reality, the beam likely encompasses multiple shock structures with various velocities and the stronger shocks are diluted within its large extent. On the other hand, the shock velocity value is not well constrained. The first way to place a constraint on this parameter would be to examine the velocity of individual clouds caught in the beam of our observations and compare it to our results.

\begin{figure}[h]
  \centering
  \includegraphics[width=\linewidth]{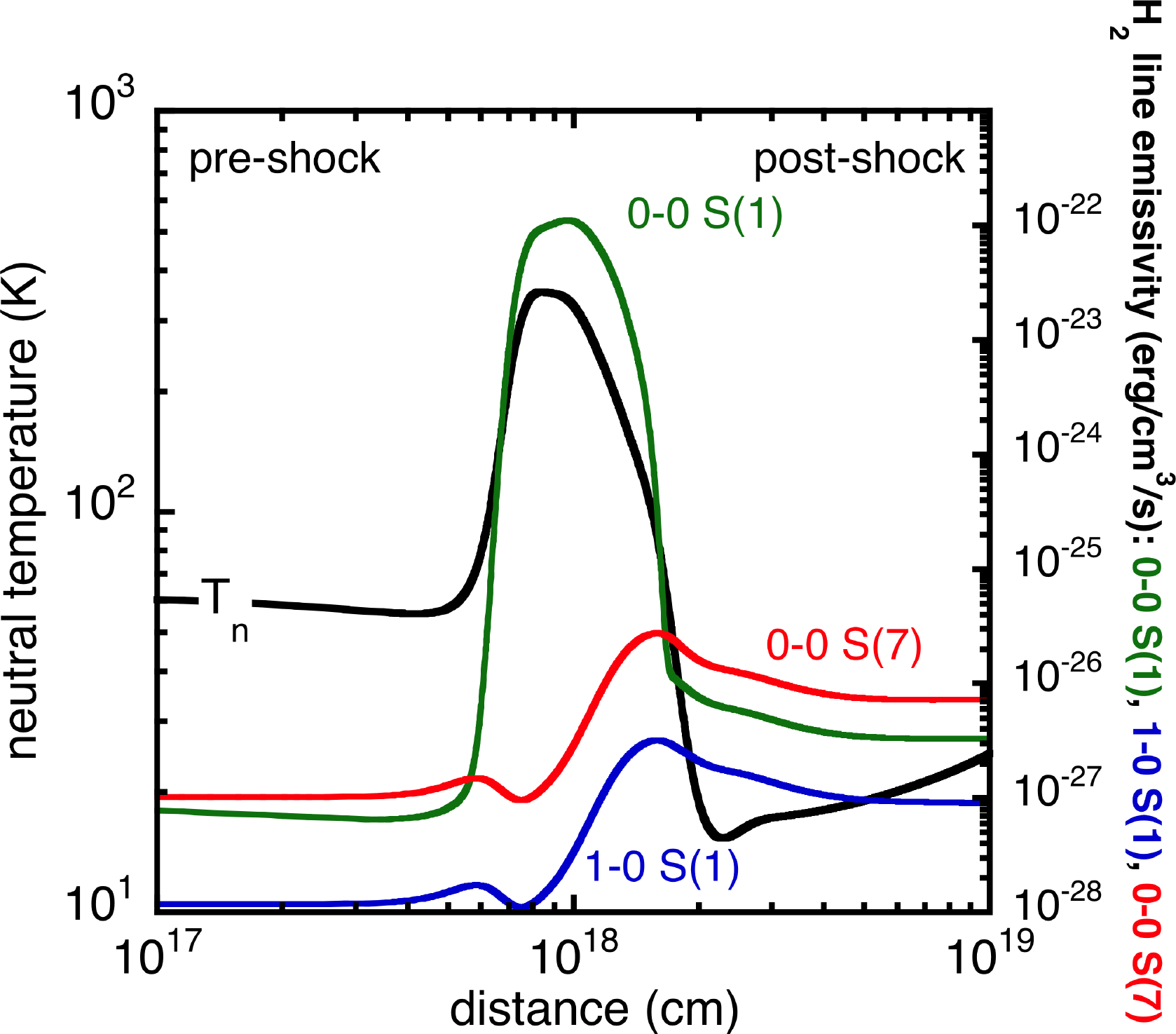}
  \caption{\label{shock_layer} Structure of the shock layer for the best-fit model with $n_H=100\: cm^{-3}$, $\varv_s=10\: km.s^{-1}$, and $G_0=0$. The black line represents the neutral temperature profile. We also show predictions of the local emissivity for three $H_2$ lines (colour lines).}
\end{figure}

   Another way to constrain the shock velocity could be to examine the ratios of the $H_2$ line integrated intensities, which are very sensitive to the shock velocity. We show the structure of the shock layer for one satisfying model with $n_H=$100 cm$^{-3}$, $G_0=0$ and a shock velocity of 10 $km.s^{-1}$ in Figure \ref{shock_layer}, which shows the evolution of the neutral temperature versus the width of the shock. In the context of comparisons with one-dimensional models considered in this work, the shock propagates along the line of sight, from the pre- to the post-shock gas (the shock is seen face on). We also show the local emissivity of three $H_2$ lines, which can be observed at high angular resolution with ground-based facilities (such as the 1-0 S(1) line at $2.12\: \mu m$), or in the near future with the \textit{James Webb Space Telescope} (such as the 0-0 S(1) and S(5) lines at respectively 17.04 and $5.51\: \mu m$). Overall, if confirmed, our result is encouraging.

   We have shown that our observations are compatible with the predictions of a shock model that is designed to compute the molecular emission from low-velocity shocks that do not have a counterpart in $H\alpha$ emission. This modelling, if confirmed, means that the shocked CO emission is caused by mechanisms internal to the clouds encompassed in the lobe of our observations. These mechanisms could be for instance related to star formation (in the form of protostellar jets and outflows), stellar winds, star deaths (supernova explosions and supernova remnants), or the natural dynamics of the interstellar matter (turbulence).

   On the other hand, with our current observations, we cannot exclude that the CO emission stems from pockets of dense molecular gas located in the post-shock of the fast jet shock, which generates $H\alpha$ emission in the same time. Other models of shocks are necessary to account for the $H\alpha$ and CO emission in this case, such as the models of high-velocity, radiative shocks shown in \cite{Hollenbach_1989}. Such radiative shocks are bright both in ionised and molecular species. Unfortunately, the grid shown in this article does not cover the appropriate conditions (in terms of pre-shock density, in particular) best suited to the description of the region. The development of similar models of fast shocks is in progress in our laboratory, from the core of the Paris-Durham shock code, so we should be able to test this assumption soon.

   In any case, we will need more observable constraints at multiple wavelengths, since the Paris-Durham shock model that we used and his spin-off in development for fast, radiative shocks are sophisticated models with a lot of input parameters. This implies that the number of observables obtained at high spatial and spectral resolution must at least match the number of input parameters, which is not quite the case yet. In particular, it is worth noting that the MUSE observations of highly ionised species presented by \cite{Santoro_2015b,Santoro_2016} and used by \cite{SalomeQ_2017} do not cover the position of the APEX pointed observations. The next observational steps are hence to obtain more observations of the molecular emission at high spatial and spectral resolution (if possible, matching the resolution of the MUSE data), and to complete the coverage of the MUSE observations to encompass the region observed with APEX.

   \subsection{Short-spacing filtering}

   Using RADEX and the Paris-Durham models, we determined that the local volume density of the molecular gas traced by the CO emission is of the order of $100\: cm^{-3}$. This is a factor of about two to three times higher than the average volume density derived geometrically in the Horseshoe complex by \cite{SalomeQ_2017}. The critical density of the CO is of the order of a few hundreds of $cm^{-3}$. Such difference in the volume density is therefore likely due to an underestimation using geometrical arguments.

   The filtering by the interferometer plays an important role in this underestimation. The combined ALMA and ACA $^{12}$CO(1-0) observations (Paper \Rmnum{2}) recover about 4.6 times the flux from ALMA only. The missing flux may result in an underestimation of the mass of the molecular clouds, and thus a lower volume density. Moreover, the majority of the molecular clouds extracted by \cite{SalomeQ_2017} have a size of about twice the ALMA beam ($1.3''\times 0.99''$). The limited spatial resolution may partially dilute the clouds and increase the observed radius. Therefore, it is probable that part of the molecular clouds are smaller than they appear. For a given mass, the larger the cloud, the lower the density. It is now essential to conduct observations with an even higher resolution.

\section{Conclusions}
\label{sec:conclusion}

   Following our previous studies in which we mapped the $^{12}$CO(2-1) emission with APEX \citep{SalomeQ_2016b} and the $^{12}$CO(1-0) emission with ALMA \citep{SalomeQ_2017} in the northern filaments of Centaurus A, we now used the Atacama Pathfinder EXperiment (APEX) to observe mid-J CO transitions in the Horseshoe complex discovered with ALMA. The $^{12}$CO(3-2) and $^{12}$CO(4-3) were detected for the first time in a region of radio jet-gas interaction, in this case between the radio jet and the H\rmnum{1} shell. Each spectrum presents two velocity components centred at velocity of about $-230$ and $-270\: km.s^{-1}$, respectively. Over the rather large APEX beam at these frequencies, the higher J transitions are not very excited on average with a SLED peaking around $J_{up}\sim 2-3$.

   In this paper, we only focussed on one region of the Horseshoe complex. Moreover, to get around the problem of the different spatial resolutions, we compared all the lines observed by APEX with the high-resolution $^{12}$CO(1-0) map from ALMA+ACA and assumed that all the clouds contained in the beam of APEX have the same physical conditions.

   We first used the non-LTE radiative transfer code RADEX to get an estimate of the physical conditions (temperature, density) of the molecular gas. A grid of models predicts a kinetic temperature of $55-70\: K$ and densities between $2-6\times 10^2\: cm^{-3}$. However, at a given kinetic temperature (resp. volume density), the blueshifted component is denser (resp. warmer) than the redshifted component.

   In addition to the $^{12}$CO(3-2) and $^{12}$CO(4-3), we also looked for dense gas tracers (HCN/$HCO^+$) and one atomic carbon line [C\rmnum{1}]. These lines were not detected, but the upper limits of the intensity line ratios with regards to $^{12}$CO(1-0) are consistent with the analysis of the CO lines, whereas they do not bring additional constrains on the physical conditions of the gas.

   We also ran a small grid of the Paris-Durham shock model. This grid tends to indicate that the pre-shock density is of the order of $100\: cm^{-3}$, corresponding to a post-shock density of a few $100\: cm^{-3}$, in agreement with the RADEX-only comparisons. The model also predicts that the region observed with APEX would have experienced a low-velocity shock, whereas the shock velocity is not well constrained. 
The results we obtained are encouraging, but more observations (higher spatial resolution, additional lines and species/tracers) will be necessary to infer the characteristics of the region more accurately.
In particular, understanding if we are witnessing shocks that originate from the inside of the clouds that we detect, or post-shock CO emission from the fast jet shock will require us to obtain additional observations, and probably to recourse to other models, that are currently in development in our laboratories.

\begin{acknowledgements}
   We thank the referee for the helpful comments. We also Carlos de Breuck and APEX operators for their support for the observations, and Miguel Querejeta for his help to combine the ALMA and ACA data. \\

   The Atacama Pathfinder EXperiment (APEX) is a collaboration between the Max-Planck-Institut für Radioastronomie, the European Southern Observatory, and the Onsala Space Observatory. \\

   PS thanks the ANR grant LYRICS (ANR-16-CE31-0011).
\end{acknowledgements}

\bibliography{Biblio}

\begin{thebibliography}{38}
\expandafter\ifx\csname natexlab\endcsname\relax\def\natexlab#1{#1}\fi

\bibitem[{Auld {et~al.}(2012)Auld, Smith, Bendo, Pohlen, Wilson, Gomez,
  Cortese, Morganti, Baes, Boselli, Cooray, Davies, Eales, Elbaz, Galametz,
  Isaak, Oosterloo, Page, Rigby, Spinoglio, \& Struve}]{Auld_2012}
Auld, R., Smith, M. W.~L., Bendo, G., {et~al.} 2012, {MNRAS}, 420, 1882

\bibitem[{Baldwin {et~al.}(1981)Baldwin, Phillips, \& Terlevich}]{Baldwin_1981}
Baldwin, J.~A., Phillips, M.~M., \& Terlevich, R. 1981, {PASP}, 93, 5

\bibitem[{Blanco {et~al.}(1975)Blanco, Graham, Lasker, \& Osmer}]{Blanco_1975}
Blanco, V.~M., Graham, J.~A., Lasker, B.~M., \& Osmer, P.~S. 1975, {ApJ}, 198,
  L63

\bibitem[{Bolatto {et~al.}(2013)Bolatto, Wolfire, \& Leroy}]{Bolatto_2013}
Bolatto, A.~D., Wolfire, M., \& Leroy, A.~K. 2013, {ARA\&A}, 51, 207

\bibitem[{Carilli \& Walter(2013)}]{Carilli_2013}
Carilli, C. \& Walter, F. 2013, {ARA\&A}, 51, 105

\bibitem[{Charmandaris {et~al.}(2000)Charmandaris, Combes, \& van~der
  Hulst}]{Charmandaris_2000}
Charmandaris, V., Combes, F., \& van~der Hulst, J.~M. 2000, {A\&A}, 356, L1

\bibitem[{Draine(1978)}]{Draine_1978}
Draine, B.~T. 1978, {ApJS}, 36, 595

\bibitem[{Flower \& Pineau~des Forêts(2015)}]{Flower_2015}
Flower, D.~R. \& Pineau~des Forêts, G. 2015, {A\&A}, 578, A63

\bibitem[{Gerin \& Phillips(2000)}]{Gerin_2000}
Gerin, M. \& Phillips, T.~G. 2000, {ApJ}, 537, 644

\bibitem[{Graham \& Price(1981)}]{Graham_1981}
Graham, J.~A. \& Price, R.~M. 1981, {ApJ}, 247, 813

\bibitem[{Gusdorf {et~al.}(2012)Gusdorf, Anderl, Güsten, Stutzki, Hübers,
  Hartogh, Heyminck, \& Okada}]{Gusdorf_2012}
Gusdorf, A., Anderl, S., Güsten, R., {et~al.} 2012, {A\&A}, 542, L19

\bibitem[{Gusdorf {et~al.}(2008)Gusdorf, Pineau~des Forêts, Cabrit, \&
  Flower}]{Gusdorf_2008}
Gusdorf, A., Pineau~des Forêts, G., Cabrit, S., \& Flower, D.~R. 2008, {A\&A},
  490, 695

\bibitem[{Harris {et~al.}(2010)Harris, Rejkuba, \& Harris}]{Harris_2010}
Harris, G. L.~H., Rejkuba, M., \& Harris, W.~E. 2010, {PASA}, 27, 457

\bibitem[{Hasegawa(1997)}]{Hasegawa_1997}
Hasegawa, T. 1997, in {IAU} Symposium, Vol. 170 ({IAU} Symposium), 39--46

\bibitem[{Hollenbach \& {McKee}(1989)}]{Hollenbach_1989}
Hollenbach, D. \& {McKee}, C.~F. 1989, {ApJ}, 342, 306

\bibitem[{Jiao {et~al.}(2017)Jiao, Zhao, Zhu, Lu, Gao, \& Zhang}]{Jiao_2017}
Jiao, Q., Zhao, Y., Zhu, M., {et~al.} 2017, {ApJL}, 840, L18

\bibitem[{Kewley {et~al.}(2006)Kewley, Groves, Kauffmann, \&
  Heckman}]{Kewley_2006}
Kewley, L.~J., Groves, B., Kauffmann, G., \& Heckman, T. 2006, {MNRAS}, 372,
  961

\bibitem[{Kim {et~al.}(2002)Kim, Martin, Stark, \& Lane}]{Kim_2002}
Kim, S., Martin, C.~L., Stark, A.~A., \& Lane, A.~P. 2002, {ApJ}, 580, 896

\bibitem[{Lee {et~al.}(2016)Lee, Madden, Lebouteiller, Gusdorf, Godard, Wu,
  Galametz, Cormier, Le~Petit, Roueff, Bron, Carlson, Chevance, Fukui,
  Galliano, Hony, Hughes, Indebetouw, Israel, Kawamura, Le~Bourlot, Lesaffre,
  Meixner, Muller, Nayak, Onishi, {Roman-Duval}, \& Sewiło}]{Lee_2016b}
Lee, M., Madden, S.~C., Lebouteiller, V., {et~al.} 2016, {A\&A}, 596, A85

\bibitem[{Leroy {et~al.}(2009)Leroy, Walter, Bigiel, Usero, Weiss, Brinks,
  de~Blok, Kennicutt, Schuster, Kramer, Wiesemeyer, \& Roussel}]{Leroy_2009}
Leroy, A.~K., Walter, F., Bigiel, F., {et~al.} 2009, {AJ}, 137, 4670

\bibitem[{Malin {et~al.}(1983)Malin, Quinn, \& Graham}]{Malin_1983}
Malin, D.~F., Quinn, P.~J., \& Graham, J.~A. 1983, {ApJ}, 272, L5

\bibitem[{{Miville-Deschênes} {et~al.}(2017){Miville-Deschênes}, Murray, \&
  Lee}]{MAMD_2017a}
{Miville-Deschênes}, M., Murray, N., \& Lee, E.~J. 2017, {ApJ}, 834, 57

\bibitem[{Morganti {et~al.}(1991)Morganti, Robinson, Fosbury,
  di~Serego~Alighieri, Tadhunter, \& Malin}]{Morganti_1991}
Morganti, R., Robinson, A., Fosbury, R. A.~E., {et~al.} 1991, {MNRAS}, 249, 91

\bibitem[{Nieten {et~al.}(1999)Nieten, Dumke, Beck, \&
  Wielebinski}]{Nieten_1999}
Nieten, C., Dumke, M., Beck, R., \& Wielebinski, R. 1999, {A\&A}, 347, L5

\bibitem[{Oka {et~al.}(1998)Oka, Hasegawa, Hayashi, Handa, \&
  Sakamoto}]{Oka_1998}
Oka, T., Hasegawa, T., Hayashi, M., Handa, T., \& Sakamoto, S. 1998, {ApJ},
  493, 730

\bibitem[{Pety {et~al.}(2017)Pety, Guzmán, Orkisz, Liszt, Gerin, Bron,
  Bardeau, Goicoechea, Gratier, Le~Petit, Levrier, Öberg, Roueff, \&
  Sievers}]{Pety_2017}
Pety, J., Guzmán, V.~V., Orkisz, J.~H., {et~al.} 2017, {A\&A}, 599, A98

\bibitem[{Rejkuba {et~al.}(2001)Rejkuba, Minniti, Silva, \&
  Bedding}]{Rejkuba_2001}
Rejkuba, M., Minniti, D., Silva, D.~R., \& Bedding, T.~R. 2001, {A\&A}, 379,
  781

\bibitem[{Sakamoto {et~al.}(1997)Sakamoto, Hasegawa, Handa, Hayashi, \&
  Oka}]{Sakamoto_1997}
Sakamoto, S., Hasegawa, T., Handa, T., Hayashi, M., \& Oka, T. 1997, {ApJ},
  486, 276

\bibitem[{Salomé {et~al.}(2016{\natexlab{a}})Salomé, Salomé, Combes, \&
  Hamer}]{SalomeQ_2016b}
Salomé, Q., Salomé, P., Combes, F., \& Hamer, S. 2016{\natexlab{a}}, {A\&A},
  595, A65

\bibitem[{Salomé {et~al.}(2016{\natexlab{b}})Salomé, Salomé, Combes, Hamer,
  \& Heywood}]{SalomeQ_2016a}
Salomé, Q., Salomé, P., Combes, F., Hamer, S., \& Heywood, I.
  2016{\natexlab{b}}, {A\&A}, 586, A45

\bibitem[{Salomé {et~al.}(2017)Salomé, Salomé, {Miville-Deschênes}, Combes,
  \& Hamer}]{SalomeQ_2017}
Salomé, Q., Salomé, P., {Miville-Deschênes}, M., Combes, F., \& Hamer, S.
  2017, {A\&A}, 608, A98

\bibitem[{Santoro {et~al.}(2016)Santoro, Oonk, Morganti, Oosterloo, \&
  Tadhunter}]{Santoro_2016}
Santoro, F., Oonk, J. B.~R., Morganti, R., Oosterloo, T.~A., \& Tadhunter, C.
  2016, {A\&A}, 590, A37

\bibitem[{Santoro {et~al.}(2015)Santoro, Oonk, Morganti, Oosterloo, \&
  Tremblay}]{Santoro_2015b}
Santoro, F., Oonk, J. B.~R., Morganti, R., Oosterloo, T.~A., \& Tremblay, G.
  2015, {A\&A}, 575, L4

\bibitem[{Schiminovich {et~al.}(1994)Schiminovich, van Gorkom, van~der Hulst,
  \& Kasow}]{Schiminovich_1994}
Schiminovich, D., van Gorkom, J.~H., van~der Hulst, J.~M., \& Kasow, S. 1994,
  {ApJ}, 423, L101

\bibitem[{Sofia {et~al.}(2004)Sofia, Lauroesch, Meyer, \&
  Cartledge}]{Sofia_2004}
Sofia, U.~J., Lauroesch, J.~T., Meyer, D.~M., \& Cartledge, S. I.~B. 2004,
  {ApJ}, 605, 272

\bibitem[{van~der Tak {et~al.}(2007)van~der Tak, Black, Schöier, Jansen, \&
  van Dishoeck}]{RADEX}
van~der Tak, F. F.~S., Black, J.~H., Schöier, F.~L., Jansen, D.~J., \& van
  Dishoeck, E.~F. 2007, {A\&A}, 468, 627

\bibitem[{Vlahakis {et~al.}(2013)Vlahakis, van~der Werf, Israel, \&
  Tilanus}]{Vlahakis_2013}
Vlahakis, C., van~der Werf, P., Israel, F.~P., \& Tilanus, R. P.~J. 2013,
  {MNRAS}, 433, 1837

\bibitem[{Weiss {et~al.}(2005)Weiss, Downes, Henkel, \& Walter}]{Weiss_2005}
Weiss, A., Downes, D., Henkel, C., \& Walter, F. 2005, {A\&A}, 429, L25

\end{thebibliography}
\bibliographystyle{aa}

\appendix
\onecolumn

\section{APEX spectra}

\begin{figure*}
  \centering
  \includegraphics[width=0.45\linewidth,trim=5 20 15 50,clip=true]{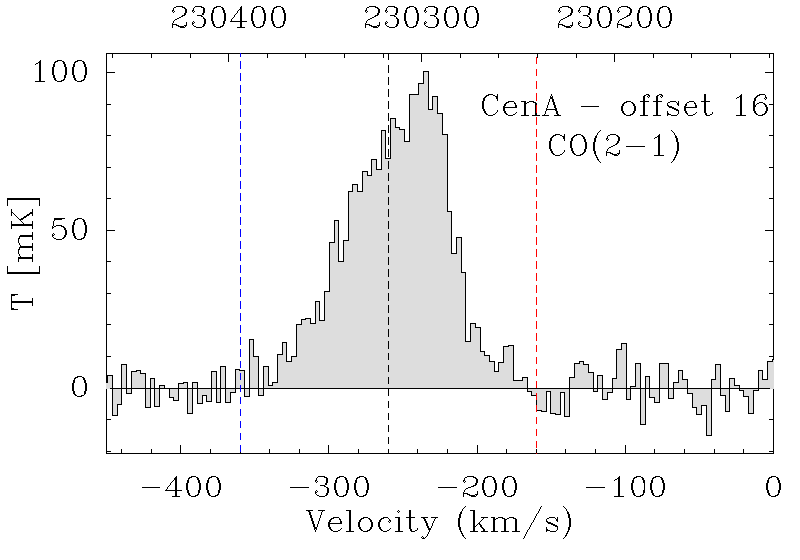}
  \hspace{3mm}
  \includegraphics[width=0.45\linewidth,trim=5 20 15 50,clip=true]{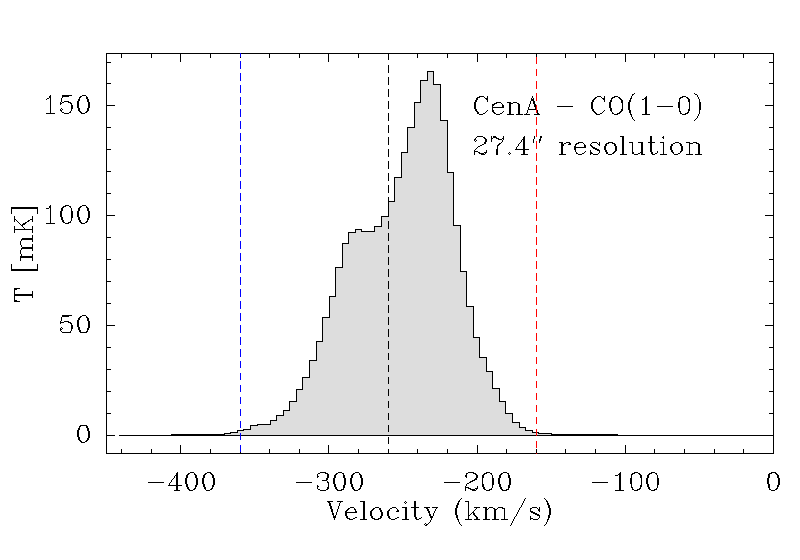} \\
  \includegraphics[width=0.45\linewidth,trim=5 20 15 50,clip=true]{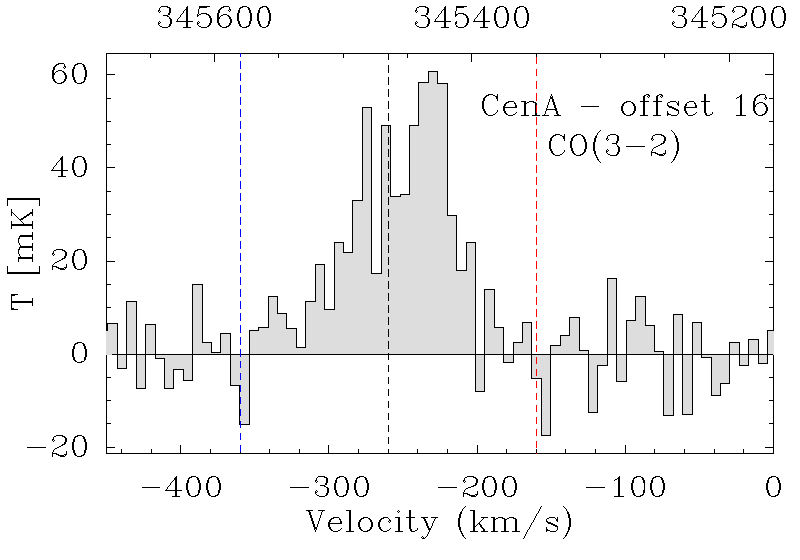}
  \hspace{3mm}
  \includegraphics[width=0.45\linewidth,trim=5 20 15 50,clip=true]{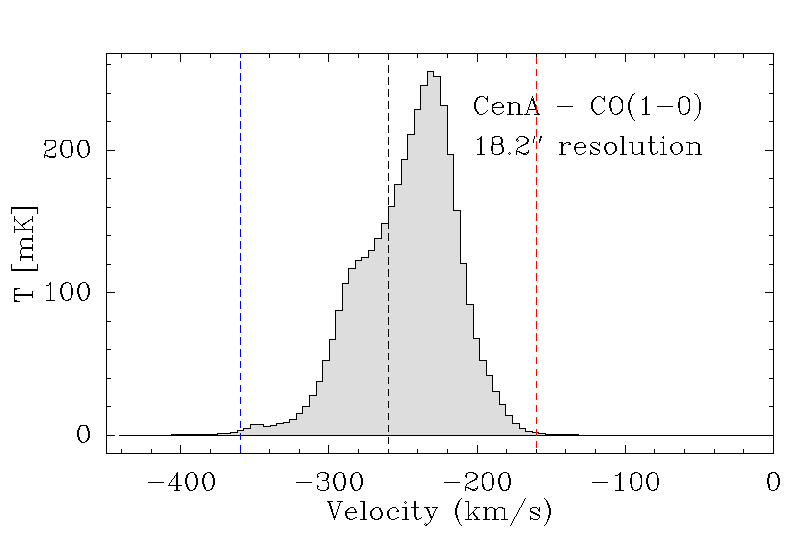} \\
  \vspace{3mm}
  \includegraphics[width=0.45\linewidth,trim=5 20 15 50,clip=true]{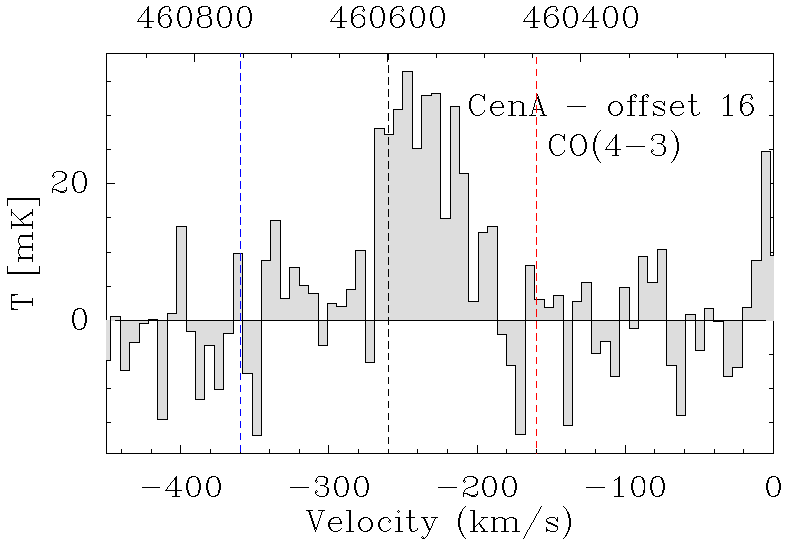}
  \hspace{3mm}
  \includegraphics[width=0.45\linewidth,trim=5 20 15 50,clip=true]{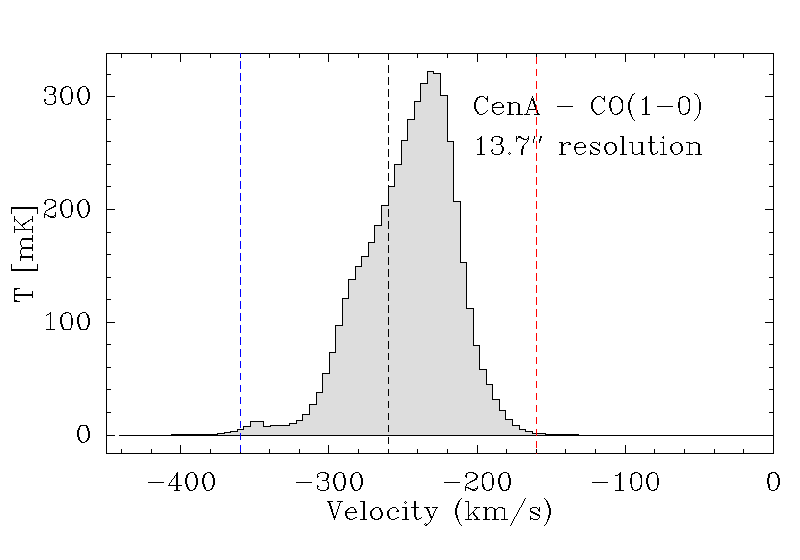}
  \caption{\label{spectra} Spectra of the $^{12}$CO(2-1), $^{12}$CO(3-2) and $^{12}$CO(4-3) emission observed with APEX (\emph{left}), and the $^{12}$CO(1-0) emission from ALMA convolved at APEX resolutions for lines shown in the left panel (\emph{right}).
For the APEX data, the spatial resolution is about $3.2\: km.s^{-1}$ for the CO(2-1) emission and $6.4\: km.s^{-1}$ for the CO(3-2) and CO(4-3) spectra. The resolution of the ALMA+ACA CO(1-0) spectra is about $4.4\: km.s^{-1}$.}
\end{figure*}

\end{document}